\documentclass[a4paper, prd,aps,tightenlines,superscriptaddress,showpacs,nofootinbib,eqsecnum, preprint]{revtex4}

\usepackage[british]{babel}
\usepackage{float}
\floatstyle{boxed}

\usepackage{graphicx,bm,dcolumn,amsmath,amsfonts,subfigure,color,latexsym,amssymb}

\def\@rcsid{\relax}
\def\rcsid#1{\def\next##1#1{\def\@rcsid{\mbox{RCS ##1}}}\next}
\def\M{{\cal M}}
\def\e{\eta}

\def\ba{\begin{eqnarray}}
\def\ea{\end{eqnarray}}
\def\be{\begin{equation}}
\def\ee{\end{equation}}

\def\nn{\nonumber}

\begin{document}


\draft

\title{Binary black hole spectroscopy}
\author{Chris Van Den Broeck}
\email{Chris.van-den-Broeck@astro.cf.ac.uk}
\author{Anand S.~Sengupta}
\email{Anand.Sengupta@astro.cf.ac.uk}

\affiliation{School of Physics and Astronomy, Cardiff University,\\
Queen's Buildings, The Parade,\\ 
Cardiff CF24 3AA, United Kingdom}

\begin{abstract}

We study parameter estimation with post-Newtonian (PN) gravitational waveforms for the quasi-circular, adiabatic inspiral of spinning binary compact objects. In particular, the performance of amplitude-corrected waveforms is compared with that of the more commonly used restricted waveforms, in Advanced LIGO and EGO. With restricted waveforms, the properties of the source can only be extracted from the phasing. In the case of amplitude-corrected waveforms, the spectrum encodes a wealth of additional information, which leads to dramatic improvements in parameter estimation. At distances of $\sim 100$ Mpc, the full PN waveforms allow for high-accuracy parameter extraction for total mass up to several hundred solar masses, while with the restricted ones the errors are steep functions of mass, and accurate parameter estimation is only possible for relatively light stellar mass binaries. At the low-mass end, the inclusion of amplitude corrections reduces the error on the time of coalescence by an order of magnitude in Advanced LIGO and a factor of 5 in EGO compared to the restricted waveforms; at higher masses these differences are much larger. The individual component masses, which are very poorly determined with restricted waveforms, become measurable with high accuracy if amplitude-corrected waveforms are used, with errors as low as a few percent in Advanced LIGO and a few tenths of a percent in EGO. The usual spin-orbit parameter $\beta$ is also poorly determined with restricted waveforms (except for low-mass systems in EGO), but the full waveforms give errors that are small compared to the largest possible value consistent with the Kerr bound. This suggests a way of finding out if one or both of the component objects violate this bound. On the other hand, we find that the spin-spin parameter $\sigma$ remains poorly determined even when the full waveform is used. Generally, all errors have but a weak dependence on the magnitudes and orientations of the spins. We also briefly discuss the effect of amplitude corrections on parameter estimation in Initial LIGO.

\end{abstract}

\pacs{04.25.Nx, 04.30.-w, 04.80.Nn, 95.55.Ym}

\maketitle

\section{Introduction}

Inspiraling compact binary objects (neutron stars and/or black holes) are among the most promising sources for existing and planned interferometric gravitational wave observatories 
\cite{Grishchuketal}. Efforts are in progress to detect their gravitational wave signals in LIGO, VIRGO, GEO600 and TAMA \cite{Observatories}.
When the gravitational wave signal enters the bandwidth of these detectors, typically the
orbits of the bodies will have approximately circularized, and the subsequent inspiral is 
described well by an adiabatic approximation in which the radii of the 
orbits are assumed not to change significantly over a single period \cite{Peters}. In this regime
the frequency and amplitude of the signal increase steadily in a ``chirp'' until 
a last stable orbit (LSO) is reached. The amplitude and phasing of signals arising from quasi-circular adiabatic inspiral have been computed to great 
accuracy using post-Newtonian (PN) methods (see \cite{Blanchet} for a review), which lead to expressions that take the form of series expansions in the orbital velocity $v$.

Apart from their obvious astrophysical importance, binary inspirals are potential ``laboratories'' for testing general relativity \cite{Testing1,Testing2,Testing3}. This is because at least in the adiabatic phase they are relatively ``clean'' and reasonably well-understood systems. However, it is important to find out how good a testing ground they offer. Precisely what information do the waveforms carry? Which parameters associated with the binary can be determined from them, and to what accuracy can this be achieved? In other words, to what extent will it be possible to reconstruct binary inspiral events from the gravitational radiation they emit?

The most accurate waveforms currently available for the quasi-circular, adiabatic inspiral of compact objects are of order $\mathcal{O}(v^5)$ in amplitude \cite{2.5PN} and $\mathcal{O}(v^7)$ in phase \cite{3.5PN}, at least in the case of non-spinning binaries; in the usual notation this corresponds to 2.5PN and 3.5PN orders, respectively. An efficient way of searching for inspiral signals in data is matched filtering \cite{Helstrom}, which involves a bank of templates. Post-Newtonian waveforms describing adiabatic inspiral are linear combinations of harmonics in the orbital phase. In most of the literature, simulated signals as well as detection templates are taken to be \emph{restricted} post-Newtonian waveforms \cite{last3minutes}, which only have the dominant harmonic at twice the orbital phase, with PN corrections to the phasing but no corrections to the amplitude. The effects on signal-to-noise ratio (SNR) and redshift reach as a result of including amplitude corrections in templates and simulating signals have been investigated in \cite{Letter,SNRpaper} in the context of both initial and advanced detectors, and it was found that the consequences are considerable. In initial detectors the use of restricted PN waveforms can lead to a significant overestimation of SNR. More interestingly, in advanced detectors the higher harmonics can give rise to considerable SNRs even when the dominant harmonic (and hence the restricted waveform) does not even enter the detector's bandwidth. As a result, the mass range of such detectors increases dramatically with the inclusion of amplitude corrections in template waveforms. Also in the case of masses for which the dominant harmonic reaches the bandwidth in only a small frequency interval, taking into account all known harmonics greatly increases the SNR. Thus, for high stellar mass to intermediate mass binaries ($100\,M_\odot \lesssim M \lesssim 1000\,M_\odot$), detection rates increase dramatically when amplitude corrections are taken into account. This also opens up the possibility of high-quality parameter estimation up to much higher masses than would be thought possible on the basis of the restricted PN approximation. 

More generally, to properly address the questions raised above, we will see that it is crucial to use the amplitude-corrected waveforms to highest available post-Newtonian order. With restricted waveforms $h_0$, almost any information one can extract about the source has to come from the phasing, i.e.~$\arg(\tilde{h}_0(f))$, with $\tilde{h}_0$ the Fourier transform of the time domain waveform. In the case of amplitude-corrected waveforms $h$, on top of the information carried by $\arg(\tilde{h}(f))$ one can exploit the intricate structure of the spectrum, i.e.~the power per frequency bin as a function of frequency. As we shall see, the additional information carried by the spectrum of the full waveform has dramatic consequences for parameter estimation, whence the title of this paper. (The study of black hole ringdown in terms of quasi-normal modes is often also referred to as ``spectroscopy'' \cite{ringdown}; in the case of inspiral the spectrum is of course continuous.)

Most of the research to date on parameter estimation with inclusion of additional harmonics has been of a preliminary nature. For Initial LIGO there is the work of Sintes and Vecchio \cite{SV1}, who used waveforms at 0.5PN order in amplitude and 2PN order in phase. Very recently, R\"over et al.~\cite{Roveretal} considered a network of initial detectors (the Initial LIGO detectors and VIRGO), taking the sources to be binary neutron star inspirals and using waveforms at 2PN in amplitude and 2.5PN in phase. For the proposed space-based detector LISA there have been studies by Sintes and Vecchio \cite{SV2} (again at 0.5PN in amplitude and 2PN in phase) and by Hellings and Moore \cite{Hellings1,Hellings2} (2PN in both amplitude and phase). None of these works took the effects of spin into account. In this paper we focus on advanced ground-based interferometric gravitational wave detectors. We will mostly consider parameter estimation with Advanced LIGO as well as a European third-generation detector that is currently under consideration \cite{Punturo}; the latter will tentatively be called the European Gravitational-wave Observatory (EGO). The waveforms used will be of 2.5PN order in amplitude and phase, with inclusion of spin-related contributions up to 2PN in the phase.

Up to 2.5PN order in amplitude, the waveforms in the two polarizations take the general form \cite{2.5PN}
\be
h_{+,\times}=\frac{2M\eta}{r} x \,
\left\{H^{(0)}_{+,\times} + x^{1/2}H^{(1/2)}_{+,\times} + x H^{(1)}_{+,\times}
+ x^{3/2}H^{(3/2)}_{+,\times} + x^2 H^{(2)}_{+,\times} + x^{5/2} H^{(5/2)}_{+,\times}
\right\} 
\label{hpluscross}
\ee
where $r$ is the distance to the binary, $M$ its total mass, and $\e$ the ratio of reduced mass to total mass. We have set $G=c=1$, as we will do throughout this paper. The post-Newtonian expansion parameter is defined as $x=(2\pi M F)^{2/3}$, with $F(t)$ the instantaneous orbital frequency. The coefficients $H^{(p/2)}_{+,\times}$, $p=0, \ldots, 5$, are linear combinations of various harmonics with prefactors that depend on the inclination angle $\iota$ of the angular momentum of the binary with respect to the line of sight as well as on $\e$; their explicit expressions can be found in \cite{2.5PN}. The measured signal also depends on the polarization angle and the position in the sky through the detector's beam pattern functions $F_{+,\times}$:
\be
h(t)=F_+ h_+(t) + F_\times h_\times(t). \label{signal}
\ee
Note that for ground-based detectors, which are the ones we will be concerned with, it is reasonable to approximate $F_{+,\times}$ as being constant over the duration of the signal. They depend on 
angles $(\theta,\phi,\psi)$, where $(\theta,\phi)$ determine sky position while $\psi$ is the polarization 
angle. The signal (\ref{signal}) is a linear combination of harmonics of the orbital phase $\Psi(t)$ with offsets $\varphi_{(k,m/2)}$:
\be
h(t) = \sum_{k=1}^N \sum_{m=0}^5 A_{(k,m/2)}(t) \cos(k\Psi(t) + \varphi_{(k,m/2)}),
\label{lincomb}
\ee
where $k$ runs over harmonics while $m/2$ is PN order in amplitude.
The coefficients $A_{(k,m/2)}$ are functions of $(r,M,\e,\theta,\phi,\psi,\iota)$ multiplied by $x^{(m+2)/2}$, with $x$ depending on time through $F(t)$. The number of harmonics $N$ depends on the PN order in amplitude; at 2.5PN one has $N=7$.

Spin-related effects will be taken into account in the following way. In the phasing, spin-orbit effects are included at 1.5PN and 2PN orders, and spin-spin effects at 2PN order, as in \cite{spin}. (The phasing formula with the inclusion of spin-orbit interactions up to 2.5PN has recently been computed \cite{spin2.5PN}, but the research we report on here was already well underway when these results appeared.) We will use the stationary phase approximation to the Fourier transform \cite{SPA}, in which the coefficients of the harmonics are proportional to the inverse square root of the time derivative of the associated instantaneous frequency. We let spin-related effects enter the amplitudes through these ``frequency sweeps'', at the same respective orders. Since we want to view the full waveform as a model for an actual signal as well as a search template, for consistency we take both amplitude and phase to be of 2.5PN order. This is a choice we make, motivated by conceptual considerations; there are no technical problems in taking the phasing up to 3.5PN.

Both the expressions for the amplitudes in (\ref{lincomb}) and the expression for the phase can be truncated to a lower post-Newtonian order. Such truncated waveforms will be
denoted as $(p,q)$PN, where $p$ refers to the PN order of the amplitude and $q$ that of the phase. Thus, the restricted PN waveform will be referred to as $(0,2.5)$PN and the
full waveform as $(2.5,2.5)$PN. 

For the Initial LIGO detector, Sintes and Vecchio \cite{SV1} discussed the improvement in error estimation gained by using amplitude-corrected waveforms by means of the covariance matrix formalism \cite{Finn}; in the present paper the same techniques will be employed. Concretely, the following questions will be addressed: (i) Using the best available waveforms, which parameters can be determined? (ii) How do the estimation accuracies improve as the PN order of the amplitude is increased? and (iii) What is the effect of the magnitudes and orientations of the components' spins?\footnote{Throughout this paper we will include a spin-orbit and a spin-spin parameter as variables in the error analysis so that even when spins are set to zero, the corresponding errors are never neglected.} 

This paper is structured as follows. In section \ref{s:Extraction} we first briefly review the covariance matrix formalism for parameter estimation. The Fourier transform of the amplitude-corrected waveforms is computed in the stationary phase approximation and its spectrum is discussed. We then apply the covariance matrix formalism to $(p,2.5)$PN waveforms with $p \geq 0.5$. In section \ref{s:Improvement} we compare the performance of restricted and amplitude-corrected waveforms in EGO and Advanced LIGO for a variety of systems. We also study the improvement in error estimation by using waveforms of order $(p,2.5)$PN as $p$ varies from zero to 2.5, and the dependence of errors on the spins of the component objects. Finally, we present results for some representative systems with Initial LIGO.
Section \ref{s:Conclusions} provides a summary and conclusions.

\section{Amplitude-corrected post-Newtonian waveforms and parameter estimation}
\label{s:Extraction}

\subsection{Overview of the covariance matrix formalism}
\label{ss:overview}

We begin by recalling some basic facts of the covariance matrix formalism for parameter estimation with matched filtering \cite{Finn}. 
To search for signals buried in the noise, the data analyst uses families of waveforms 
$\tilde{h}[\theta^a](f)$, determined by a 
finite number of parameters $\theta^a$ that characterize the binary. 
The measured values of the parameters, $\hat{\theta}^a$, depend on the realization of the detector noise 
at the time the signal was present, so that in general they do not correspond to the 
``true'' values $\bar{\theta}^a$. For large
SNR, the measurement errors $\Delta \theta^a=\bar{\theta}^a-\hat{\theta}^a$ follow 
a multivariate Gaussian probability distribution \cite{Finn},
\be
P(\Delta \theta^a) = P_0
       \exp(-\Gamma_{bc}(\hat{\theta}^a) \Delta \theta^b \Delta \theta^c/2)
\label{distribution}
\ee
with $P_0$ a normalization factor, and $\Gamma_{ab}(\theta^a)$ is the Fisher information matrix. The latter may be thought of as a metric on the differentiable manifold naturally defined by the family of waveforms 
$\{h[\theta^a]\}$, and it is given by
\be
\Gamma_{ab} = ( h_a | h_b ),
\label{Fisher}
\ee
where $h_a \equiv \partial h/\partial \theta^a$, and $( \, .\, | \, . )$
is the usual product defined by the one-sided noise power spectral density $S_h(f)$: 
\be
( x | y ) \equiv 4 \int_{f_{min}}^{f_{max}} df\,\frac{\mbox{Re}(\tilde{x}^\ast(f) \tilde{y}(f))}{S_h(f)}.
\label{product}
\ee
The integration domain $[f_{min},f_{max}]$ is determined both by the detector and by the nature of the signal. As explained in \cite{SNRpaper} (and in subsection \ref{ss:waveforms} below), it is reasonable to cut off the $k$th harmonic in the waveform at a frequency $k f_{LSO}$, where $f_{LSO}$ may be viewed as the orbital frequency at the last stable orbit. Since the amplitude-corrected waveform contains 7 harmonics, in that case we will effectively have $f_{max} = 7 f_{LSO}$, while for the restricted waveform $f_{max} = 2 f_{LSO}$. As to the lower cut-off frequency, power spectral densities $S_h(f)$ tend to rise very quickly below a certain frequency $f_s$ where they can be considered infinite for all practical purposes; accordingly, we take $f_{min} = f_s$.

The signal-to-noise ratio is also defined in terms of the product (\ref{product}),
\be
\rho[h] \equiv \sqrt{(h|h)}.
\ee

The inverse of the information matrix is the covariance matrix $\Sigma^{ab}$, and one has
\be
\Sigma^{ab} \equiv (\Gamma^{-1})^{ab} = \langle \Delta\theta^a \Delta \theta^b \rangle
\ee
where $\langle\,.\,\rangle$ denotes the average with respect to the probability distribution $P(\Delta\theta^a)$
(\ref{distribution}). The root-mean-square error in the measurement of the parameters is then
\be
\sigma^a \equiv \sqrt{\langle (\Delta\theta^a)^2 \rangle} = \sqrt{\Sigma^{aa}},
\label{RMS}
\ee
and with slight abuse of notation these will simply be denoted $\Delta \theta^a$ in the rest of the paper.
Finally, the correlation coefficients are defined by
\be
c^{ab} \equiv \frac{\Sigma^{ab}}{\sqrt{\Sigma^{aa}\Sigma^{bb}}}.
\label{corrcoeffs}
\ee
(Note that repeated indices in Eqs.~(\ref{RMS}) and (\ref{corrcoeffs}) are
not meant to be summed over.) By definition one has $-1 \leq c_{ab} \leq 1$. If 
the correlation coefficient between two parameters is close to 1 (resp.~$-1$), they are 
strongly correlated (resp.~anticorrelated), meaning that one of them is redundant. A correlation coefficient being close to zero indicates that the parameters are largely unrelated. 

Usually the Fisher information matrix, and hence errors and correlations, can not be calculated exactly because the integrals defining $\Gamma_{ab}$ can not be performed analytically. This is also the case here, and most of the results in this paper will refer to numerical calculations with the software package \emph{Mathematica}. The matrix $\Sigma = \Gamma^{-1}$ was obtained using the built-in matrix inversion routine. In that regard it is worth mentioning that problems can arise when numerically inverting large matrices such as the Fisher matrices encountered here; we will come back to this point in subsection \ref{ss:parameterestimation} below. 

The covariance formalism is only valid for sufficiently high SNR; at low SNR it tends to underestimate errors, as seen in Monte Carlo treatments involving large numbers of simulations with different realizations of the noise \cite{MonteCarlo}. However, at distances of a few hundred Mpc the SNRs in Advanced LIGO and EGO will be rather high, as was made evident in \cite{SNRpaper}; see also subsection \ref{ss:SNRs} below.

\subsection{Amplitude-corrected PN waveforms}
\label{ss:waveforms}

The Fisher matrix (\ref{Fisher}) is defined in terms of (derivatives of) the Fourier transform of the waveform 
$h[\theta^a](t)$. We will find it convenient to use the well-known stationary phase approximation (SPA) \cite{SPA} to the Fourier transform in order to compute the scalar products involved. The resulting waveforms have already been discussed in detail in \cite{SNRpaper}; here we only give a quick overview. 

As mentioned in the Introduction, the waveforms $h[\theta^a](t)$ take the form
\be
h(t) = \sum_{k=1}^7 h^{(k)}(t) \label{sum}
\ee
where the $h^{(k)}(t)$ involve harmonics of the orbital phase with constant offsets: 
\be
h^{(k)}(t) = \sum_{m=0}^5 A_{(k,m/2)}(t) \cos(k\Psi(t) + \varphi_{(k,m/2)}). \label{harmonic}
\ee  
These harmonics are found from (\ref{hpluscross}) and (\ref{signal}); sky position $(\theta,\phi)$ and the polarization angle $\psi$ enter the coefficients $A_{(k,m/2)}$ and offsets $\varphi_{(k,m/2)}$ through the beam pattern functions $F_{+,\times}$:
\ba
F_+(\theta,\phi,\psi) &=&  \frac{1}{2}\left(1+\cos^2(\theta)\right)\cos(2\phi)\cos(2\psi) 
- \cos(\theta)\sin(2\phi)\sin(2\psi), \nn\\
F_\times(\theta,\phi,\psi) &=& \frac{1}{2}\left(1+\cos^2(\theta)\right)\cos(2\phi)\sin(2\psi) 
+ \cos(\theta)\sin(2\phi)\cos(2\psi),
\label{beampatternfunctions}
\ea
which in the case of ground-based detectors can be considered constant for the duration of the observed part of the signal.

During the inspiral phase one has $|d\ln A_{(k,s)}/dt| \ll 1$ and 
$|k d^2\Psi/dt^2| \ll (k d\Psi/dt)^2$, in which case we can use the stationary phase approximation \cite{SPA} to the Fourier transform of (\ref{harmonic}). For positive frequencies we have
\ba
\tilde{h}^{(k)}(f) &\simeq& \frac{\sum_{m=0}^5 A_{(k,m/2)}\left(t(\frac{1}{k}f)\right)}{2\sqrt{k\dot{F}\left(t\left(\frac{1}{k}f\right)\right)}}
                \exp\left[i\left(2\pi f t\left(\frac{1}{k}f\right) - k\Psi\left(t\left(\frac{1}{k}f\right)\right) - \varphi_{(k,m/2)} - \pi/4\right)\right] \nn\\
               &=& \frac{\sum_{m=0}^5 A_{(k,m/2)}\left(t(\frac{1}{k}f)\right)\,e^{-i\varphi_{(k,m/2)}}}{2 \sqrt{k\dot{F}\left(t(\frac{1}{k}f)\right)}} 
\exp\left[i\left(2\pi f t_c -\pi/4 + k \psi\left(\frac{1}{k} f\right)\right)\right].  \label{h_k} 
\ea
A dot denotes derivation with respect to time and $t_c$ is the coalescence time. 

We recall that the function $F(t)$ is the instantaneous orbital frequency. The function $t(f)$ is defined implicitly by
$F(t(f)) = f$. In the expressions (\ref{h_k}) for the $\tilde{h}^{(k)}$ we take the ``frequency sweep'' to be \cite{2.5PN,3.5PN,Arunetal}
\ba
\dot{F} &=& \frac{96}{5\pi\M^2}(2 \pi \M F)^{11/3} \left[1 - \left(\frac{743}{336}+\frac{11}{4}\eta\right)(2 \pi M F)^{2/3} + (4\pi - \beta)(2 \pi M F)  \right. \nn\\
&& \left. \quad + \left(\frac{34103}{18144}+\frac{13661}{2016}\eta + \frac{59}{18}\eta^2 + \sigma\right)(2 \pi M F)^{4/3} - \left(\frac{4159\pi}{672} + \frac{189\pi}{8}\eta\right)(2 \pi M F)^{5/3}\right],
\nn\\
\label{freqsweep}
\ea
where $\M=M\e^{3/5}$ is the chirp mass. The parameter $\beta$ occurring at 1.5PN order encapsulates the leading-order spin-orbit effects, while the parameter $\sigma$ at 2PN order also encodes spin-spin effects; at 2.5PN order we neglect all spin-related effects. $\beta$ and $\sigma$ are given by \cite{KWW}:
\ba
\beta &=& \frac{1}{12} \sum_{i=1}^2 \left[113\,(m_i/M)^2 + 75 \,\e \right] \,\hat{L} \cdot \bar{\chi}_i,\label{beta}\\
\sigma &=& \frac{\eta}{48}\left[-247 \,(\bar{\chi}_1 \cdot \bar{\chi}_2) + 721 \,(\hat{L} \cdot \bar{\chi}_1)(\hat{L} \cdot \bar{\chi}_2)\right],\label{sigma}
\ea
with $\bar{\chi}_i = \vec{S}_i/m_i^2$ and $\vec{S}_i$, $i=1,2$, the spins of the binary's components, while $\hat{L}$ is the unit vector in the direction of orbital angular momentum.
To a reasonable approximation, $\beta$ and $\sigma$ may be considered constant \cite{CF}. The maximum possible value of $|\sigma|$ for sub-extremal black holes is $79/32 \simeq 2.5$ and the maximum value of $|\beta|$ is $113/12 \simeq 9.4$.

To 2.5PN order, the phase $\psi(f)$ is given by
\be
\psi(f) = -\psi_c + \frac{3}{256\,(\pi \M f)^{5/3}}\sum_{i=0}^5 \psi_i (2 \pi M f)^{i/3},
\ee 
where $\psi_c$ is the orbital phase at coalescence. We take the coefficients $\psi_i$ to be \cite{2.5PN,3.5PN,Arunetal}
\ba
\psi_0 &=& 1,\nn\\
\psi_1 &=& 0,\nn\\
\psi_2 &=& \frac{20}{9}\,\left[\frac{743}{336} + \frac{11}{4}\eta\right],\nn\\
\psi_3 &=& -4(4\pi - \beta),\nn\\
\psi_4 &=& 10\,\left[\frac{3058673}{1016064} + \frac{5429}{1008}\eta + \frac{617}{144}\eta^2 - \sigma \right],\nn\\
\psi_5 &=& \pi\,\left[\frac{38645}{756} + \frac{38645}{756}\ln\left(\frac{f}{f_{LSO}}\right) - \frac{65}{9}\eta\left(1 + \ln\left(\frac{f}{f_{LSO}}\right)\right)\right].  
\ea

The SPA for the $(p,2.5)$PN
waveform is then
\be
\tilde{h}_{SPA}(f) = \sum_{k=1}^{N_p} \left[\frac{\sum_{m=0}^{2p} A_{(k,m/2)}\left(t\left(\frac{1}{k}f\right)\right)\,e^{-i\varphi_{(k,m/2)}}}{2\sqrt{k \dot{F}\left(t\left(\frac{1}{k} f\right)\right)}}\right]_p 
\exp\left[i\left(2\pi f t_c -\pi/4 + k \psi\left(\frac{1}{k} f\right)\right)\right], 
\label{SPA}
\ee
where $N_p$ is the number of harmonics, and $[\,.\,]_p$ denotes consistent truncation to $p$th post-Newtonian order (i.e., the ``Newtonian'' prefactor $f^{-7/6}$ is taken outside and the remaining expression is Taylor-expanded in $(2\pi M f)^{1/3}$ up to order $(2\pi M f)^{5/3}$). For more explicit expressions we refer to \cite{SNRpaper}.

In the time domain it is reasonable to cut off the waveform at a time corresponding to $F(t) = f_{LSO}$, where $f_{LSO}$ is the orbital frequency at ``last stable orbit''. For simplicity we use the expression corresponding to the extreme mass ratio:
\be
f_{LSO} = \frac{1}{6^{3/2} 2\pi M},
\ee
where $M$ is the total mass of the system. In the frequency domain this roughly corresponds to terminating the harmonics at multiples of $f_{LSO}$; in practice we multiply the truncated $k$th harmonic by $\theta(k f_{LSO} - f)$ where $\theta(x)$ is the usual Heaviside function ($\theta(x) = 1$ if $x > 0$ and $\theta(x)=0$ otherwise) \cite{SNRpaper}.

Finally, in \cite{FFT1,FFT2} the stationary phase approximation was compared with an alternative way of approximating the Fourier transform, the so-called fast Fourier transform (FFT), for initial detectors and relatively small masses. In that case the SPA and FFT differ by only a few percent, and the situation improves dramatically with increased detector sensitivity at low frequencies. We also note that the conditions for the applicability of the SPA as spelled out above are satisfied more easily with increasing $k$.

\subsection{Spectrum as observed in a detector}
\label{ss:structure}

Our goal is to compare the quality of parameter estimation with restricted and amplitude-corrected waveforms. As we now explain, the extent to which the full waveform will perform better is determined by the information content of its spectrum. Consider the contribution to the SNR squared per logarithmic frequency bin, as a function of frequency. The squared signal-to-noise ratio of a waveform $h$ in a detector with noise power spectral density $S_h(f)$ is
\ba
\rho^2[h] &=& 4 \int_{f_{min}}^{f_{max}} \frac{|h(f)|^2}{S_h(f)}df \nn\\
&=& 4 \int_{f_{min}}^{f_{max}} \frac{f\,|h(f)|^2}{S_h(f)} d\ln(f).
\ea
We now define the ``observed spectrum'' as
\be
\mathcal{P}(f) \equiv \frac{f\,|h(f)|^2}{S_h(f)}.
\label{ObservedSpectrum}
\ee
The name is apt; the observed spectrum bears a direct relationship to the way signals are seen in a detector, which depends both on the waveform and on the sensitivity of the detector. In \cite{FFT2} the quantity $\mathcal{P}(f)$ was used (under another name) as a diagnostic to compare different (restricted) waveform approximants. 

When using restricted waveforms $h_0$ for parameter estimation, of necessity almost all information one can extract about the source has to come from the phasing, $\arg(\tilde{h}_0(f))$; the spectrum does not play much of a role. The situation is very different when amplitude-corrected waveforms $h$ are used. In that case we can exploit not only $\mbox{arg}(\tilde{h}(f))$; on top of that there is the wealth of information contained in $|\tilde{h}(f)|^2$, or more directly in the observed spectrum $\mathcal{P}(f)$.

\begin{figure}[htbp!]
\centering
\includegraphics[scale=0.40,angle=0]{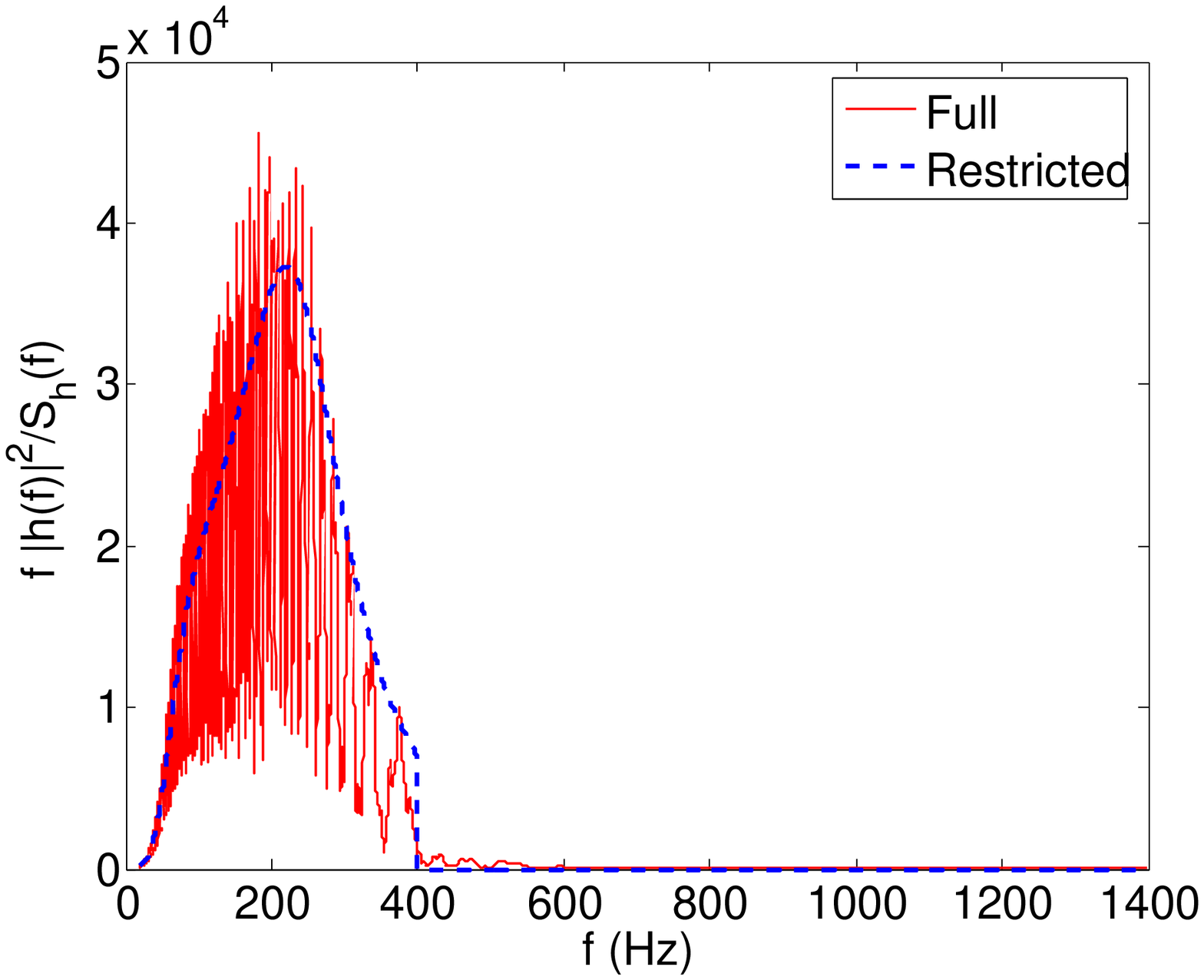}
\includegraphics[scale=0.40,angle=0]{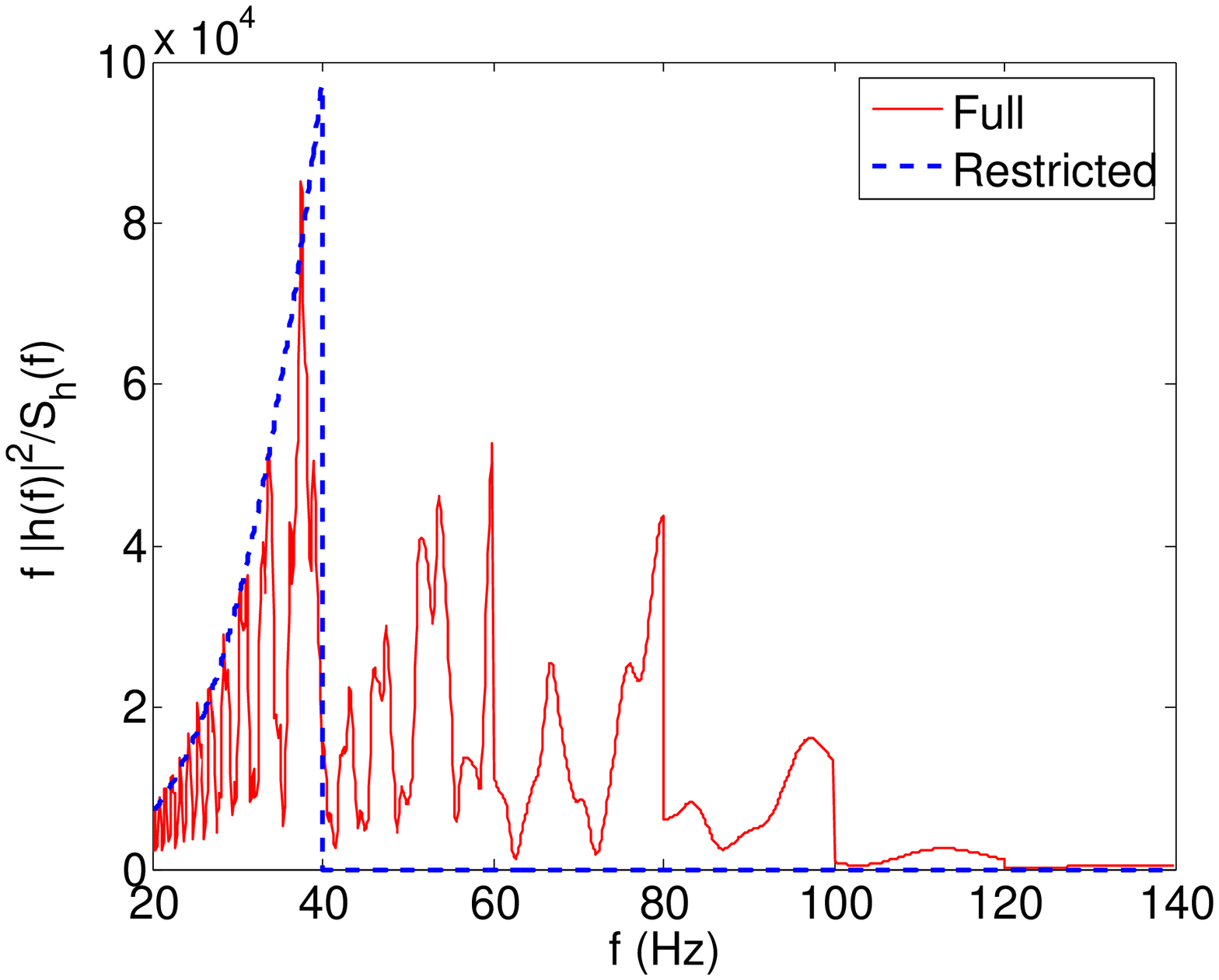}
\caption{The observed spectrum in Advanced LIGO for a $(1,10)\,M_\odot$ system (left) and a $(10,100)\,M_\odot$ system (right). Distance is 100 Mpc and angles were chosen arbitrarily as $\theta=\phi=\pi/6$, $\psi=\pi/4$, $\iota=\pi/3$. Spins are zero.}
\label{f:spectra}
\end{figure} 

In Fig.~\ref{f:spectra} we have plotted $\mathcal{P}(f)$ for both restricted and amplitude-corrected waveforms, for two different systems in Advanced LIGO. With a $(1,10)\,M_\odot$ system, the restricted waveform $h_0$ will penetrate a detector's bandwidth up to a relatively high frequency, as $2 f_{LSO} = 1/(6^{3/2}\pi M)$ will be large. The higher harmonics that are present in the full waveform continue to still higher frequencies, but their contribution beyond $2 f_{LSO}$ is suppressed by the denominator $S_h(f)$ in (\ref{ObservedSpectrum}); Advanced LIGO has its peak sensitivity at about 215 Hz. On the other hand, for a $(10,100)\,M_\odot$ system, $h_0$ will deposit power in a much smaller frequency interval, with an upper cut-off frequency much below the frequency where the detector is most sensitive. Therefore, for the restricted waveform, the quantity $\mathcal{P}(f)$ will be non-zero in only a small interval in which the denominator $S_h(f)$ in (\ref{ObservedSpectrum}) is large. By contrast, the higher harmonics in the amplitude-corrected waveform $h$ can penetrate the bandwidth up to frequencies where the detector is much more sensitive; note the distinctive features at 40 Hz, 60 Hz, 80 Hz, $\ldots$.

For both systems, $\mathcal{P}(f)$ for the full waveform exhibits a great deal of structure due to interference between the various harmonics.
As a consequence, we may expect the amplitude-corrected waveforms to allow for parameter extraction with significantly smaller errors than the restricted waveforms, with the difference being much more pronounced for higher masses.

\subsection{Parameter estimation with amplitude-corrected waveforms: Overview} 
\label{ss:parameterestimation}

Before going into details, we begin with an overview and discuss parameter estimation with the waveforms of the previous subsection in a qualitative way.

For \emph{any} family of waveforms of order $(p,q)$PN with $0.5 \leq p < \infty$, the metric $\Gamma_{ab}$
has a coordinate singularity at $\e=1/4$. This can readily be inferred from the explicit expression for $(0.5,q)$PN waveforms in \cite{SV1}: $h_{\ln(\e)}$ blows up as 
$\eta \rightarrow 1/4$ and the same is true of $\Gamma_{\ln(\e)\,b} = (h_{\ln(\e)}| 
h_b )$. The problem does not occur if the usual coordinate $\ln(\e)$ is replaced by
$\delta$, where
\be
\delta \equiv \left|\frac{m_1-m_2}{m_1 + m_2}\right|.
\ee
(This new parameter was already introduced in \cite{Hellings1}, but there the motivation was left unclear.)
For this reason, when considering amplitude-corrected waveforms we will use the set of parameters
\be
(\theta^a) = (\ln(1/r),\ln(\M),\delta,t_c,\psi_c,\cos(\iota),\cos(\theta),\phi,\psi,\beta,\sigma).
\label{coordinates}
\ee

Generally, not all parameters affecting the waveform can be extracted a posteriori, and this will also be the case here. As the PN order in amplitude is increased, one has the following:

\begin{itemize}

\item The restricted ($p=0$) waveform, whose SPA takes the form 
\be
\mathcal{A} f^{-7/6}\,\exp\left[ i \left(2\pi f t_c -\pi/4 + 2 \psi(f/2)\right) \right],
\ee
only allows for the determination of \cite{Arunetal,PoissonWill}
\begin{equation}
(\ln(\mathcal{A}),\ln(\M),\delta,t_c,\psi_c,\beta,\sigma),
\label{restrictedparameters}
\end{equation}
where the amplitude $\mathcal{A}$ depends on $r$, $\M$, $\delta$, and the four angular parameters. 

\item Going to 0.5PN in amplitude, two additional harmonics appear and it becomes possible \emph{in principle} to discriminate between three variables appearing in the amplitudes: The distance $r$, the inclination angle $\iota$, and e.g.~the quotient $F_\times/F_+$ of the beam pattern functions \cite{SV1}. Below we will instead include $(\ln(1/r),\cos(\iota),\cos(\theta))$, as these parameters have a direct physical meaning.

\item At 1PN in amplitude there are four harmonics in all so that one might expect either $\phi$ or $\psi$ to be an additional measurable parameter. However, whichever one chooses, the errors on $(\cos(\theta),\phi,\psi)$ tend to be extremely large and any of these parameters will be unmeasurable in practice. The same goes for the distance $r$. This remains true up to 2.5PN order in amplitude. 

\end{itemize}

Ideally one would like to work with the full $11\,\times\,11$ Fisher matrix based on all of the variables in (\ref{coordinates}). However, inversion of such a large matrix tends to fail when using the built-in inversion routine of \emph{Mathematica}. The reason is that the matrix will often be ``ill-conditioned''; this problem was also encountered in \cite{Testing1}, and a good discussion can be found in Appendix B of that paper. We were able to avoid this problem by instead considering a Fisher matrix that takes into account a smaller number of variables but leads to accurate estimates for the errors on the parameters one would be the most interested in. As discussed above, all variables in (\ref{coordinates}) except for $r$ and $(\cos(\theta),\phi,\psi)$ can be determined at the highest post-Newtonian order considered in this paper. As we shall see, the latter three are strongly (anti-)correlated among each other and have only tiny correlations with all other parameters. On the basis of this one might conclude that it is safe to disregard them entirely and build an $8\,\times\,8$ Fisher matrix associated with the eight remaining parameters,
\be
(\ln(1/r),\ln(\M),\delta,t_c,\psi_c,\cos(\iota),\beta,\sigma). \label{8parameters}
\ee
By comparing results for a representative sample of points in parameter space where inversion of both the $8\,\times\,8$ and $11\,\times\,11$ Fisher matrices did succeed, we were able to ascertain that it is indeed safe to use the smaller matrix, except for the following important caveat. Use of the $8\,\times\,8$ Fisher matrix leads to only a few percent difference in the errors on
\be
(\ln(\M),\delta,t_c,\cos(\iota),\beta,\sigma), \label{robust}
\ee
but it leads to a deceptively small error on distance ($\Delta r/r \sim 0.1$ for a large range of masses in Advanced LIGO and about ten times smaller in EGO). The reason is that excluding $(\cos(\theta),\phi,\psi)$ from the analysis is equivalent to assuming that they are known, fixed parameters. Since $\M^{5/6}/r$ is an overall prefactor in the waveform and because $\M$ is already well-determined from the phase alone \cite{Arunetal,PoissonWill}, the error on distance then also comes out to be small. In reality, with inclusion in the analysis of sky position and polarization angle, the relative error on distance tends to be far larger than 100\%. This is as one would expect for short-lived sources in a single interferometer \cite{Tinto}. Unless stated otherwise we will work with the $8\,\times\,8$ Fisher matrix corresponding to the eight variables (\ref{8parameters}), which leads to a robust estimation of the parameters (\ref{robust}), and the values for the errors on $\ln(1/r)$ and $\psi_c$ will be disregarded. For most choices of parameter values, this smaller Fisher matrix is free of the abovementioned problem with matrix inversion. 

For \emph{restricted} waveforms, the $6\,\times\,6$ Fisher matrix based on the last six parameters\footnote{The Fisher matrix based on \emph{all} of the parameters (\ref{restrictedparameters}) will be block-diagonal, and the block associated with the variable $\ln(\mathcal{A})$ can be ignored \cite{Arunetal}.} in (\ref{restrictedparameters}) becomes ill-conditioned when $m_1/m_2 \simeq 1$. The reason is that, unlike with the full waveforms, with restricted waveforms the error $\Delta \delta$ grows very quickly as $m_1/m_2 \rightarrow 1$. To understand how this comes about, consider a coordinate system in which $\ln(\eta)$ is used in place of $\delta$, as in \cite{Arunetal}. One has $\Delta\eta/\eta = |d \ln(\eta)/d \delta|\,\Delta\delta$, and $|d\ln(\eta)/d\delta| = 2\delta/(1-\delta^2) \rightarrow 0$ as $m_1/m_2 \rightarrow 1$, while $\Delta\eta/\eta$ remains finite; hence $\Delta\delta \rightarrow \infty$. The result is that in our framework, we will not be able to compare parameter estimation with restricted versus amplitude-corrected waveforms when $m_1/m_2 \simeq 1$. Indeed, in that case the Fisher matrix for \emph{restricted} waveforms is ill-conditioned if $\delta$ is used as a coordinate, while the Fisher matrix for the \emph{full} waveforms becomes singular at $m_1 = m_2$ if $\ln(\eta)$ is used in place of $\delta$.

With amplitude-corrected waveforms, the mass-related parameters $\M$ and $\delta$ and also the coalescence time $t_c$ can be measured with great accuracy for a wide range of systems. One tends to have $\Delta \cos(\iota) \sim 0.2$ with Advanced LIGO and ten times smaller with EGO. However, in the absence of information on the polarization angle the information carried by the inclination angle is not very useful, and we will not put much emphasis on it.

Finally, a comment on the spin-related parameters $\beta$ and $\sigma$. In previous works \cite{CF,PoissonWill} one would multiply the error distribution (\ref{distribution}) by Gaussians in $\beta$ and $\sigma$ with appropriate widths to reflect the fact that the absolute values of these parameters have very specific upper bounds if the components of the binary are sub-extremal Kerr black holes. In the context of restricted PN waveforms it was shown in \cite{PoissonWill} that doing so has the effect of decreasing error estimates for $(\ln(\M),\ln(\eta),t_c,\psi_c)$. However, apart from the fact that the method is somewhat inelegant, it also inculcates our biases as to what the endpoint of gravitational collapse should be. Gravitational wave astronomy will provide us with an opportunity to directly search for objects whose spins exceed the Kerr bound, as we will discuss below. Accordingly we will not put any a priori constraints on the values of $\beta$ and $\sigma$. The parameter $\beta$ is well determined in both Advanced LIGO and EGO; $\sigma$ is unmeasurable in Advanced LIGO and it is rather poorly determined also in EGO.

In summary, even with $(2.5,2.5)$PN waveforms the number of parameters that can be determined with reasonable accuracy is the same as for the restricted ones. With restricted waveforms the well-determined variables are $(\ln(\mathcal{A}),\ln(\M),\delta,t_c,\beta,\sigma)$; with the $(2.5,2.5)$PN waveforms they are $(\ln(\M),\delta,t_c,\cos(\iota),\beta,\sigma)$. 

\section{Reconstruction of binary inspiral events using advanced detectors}
\label{s:Improvement}

We now study the quality of parameter estimation to be expected with Advanced LIGO and EGO, comparing the performance of restricted and amplitude-corrected waveforms for two sets of binary systems. In one set, the mass ratio $q_m = m_1/m_2$ is held fixed while the total mass $M$ is varied; in the other, $M$ is kept the same but $q_m$ is varied. The performance of restricted PN waveforms will be compared with the more general, $(2.5,2.5)$PN ones. We will mostly be interested in asymmetric systems where one component is much heavier than the other since it is for such systems that we expect the higher-order amplitude corrections to make the biggest difference.  When varying $M$ for fixed $q_m$, for the most part we restrict attention to systems with total mass up to $\sim 250\,M_\odot$ in Advanced LIGO and $\sim 500\,M_\odot$ in EGO. (Moderately heavy and moderately asymmetric systems with masses in the order of hundreds of solar masses are likely to be formed in galactic nuclei in the process that leads to the formation of supermassive black holes, or in globular clusters \cite{QuinlanShapiro}.) We will also look at the dependence of errors on the PN order in amplitude. Initially the spins of the binary's components will be set to zero so that $\beta = \sigma = 0$; near the end of this section we investigate how errors depend on the values of the spins.

The set of parameters (\ref{coordinates}) can naturally be divided into two subsets:
\begin{quote}

(i) The ``extrinsic'' parameters $(t_c,\psi_c,\ln(1/r),\cos(\theta),\phi,\cos(\iota),\psi)$ which describe the time at which coalescence occurred, the phase at coalescence, the distance, the sky position, and the orientation of the inspiraling binary with respect to the detector;

(ii) The ``intrinsic'' parameters $(\ln(\M),\delta,\beta,\sigma)$ which encode the masses and spins of the binary's components.

\end{quote}

We discuss in turn the behavior of signal-to-noise ratios, errors on extrinsic parameters, and errors on intrinsic parameters.

\begin{figure}[htbp!]
\centering
\includegraphics[scale=0.40,angle=0]{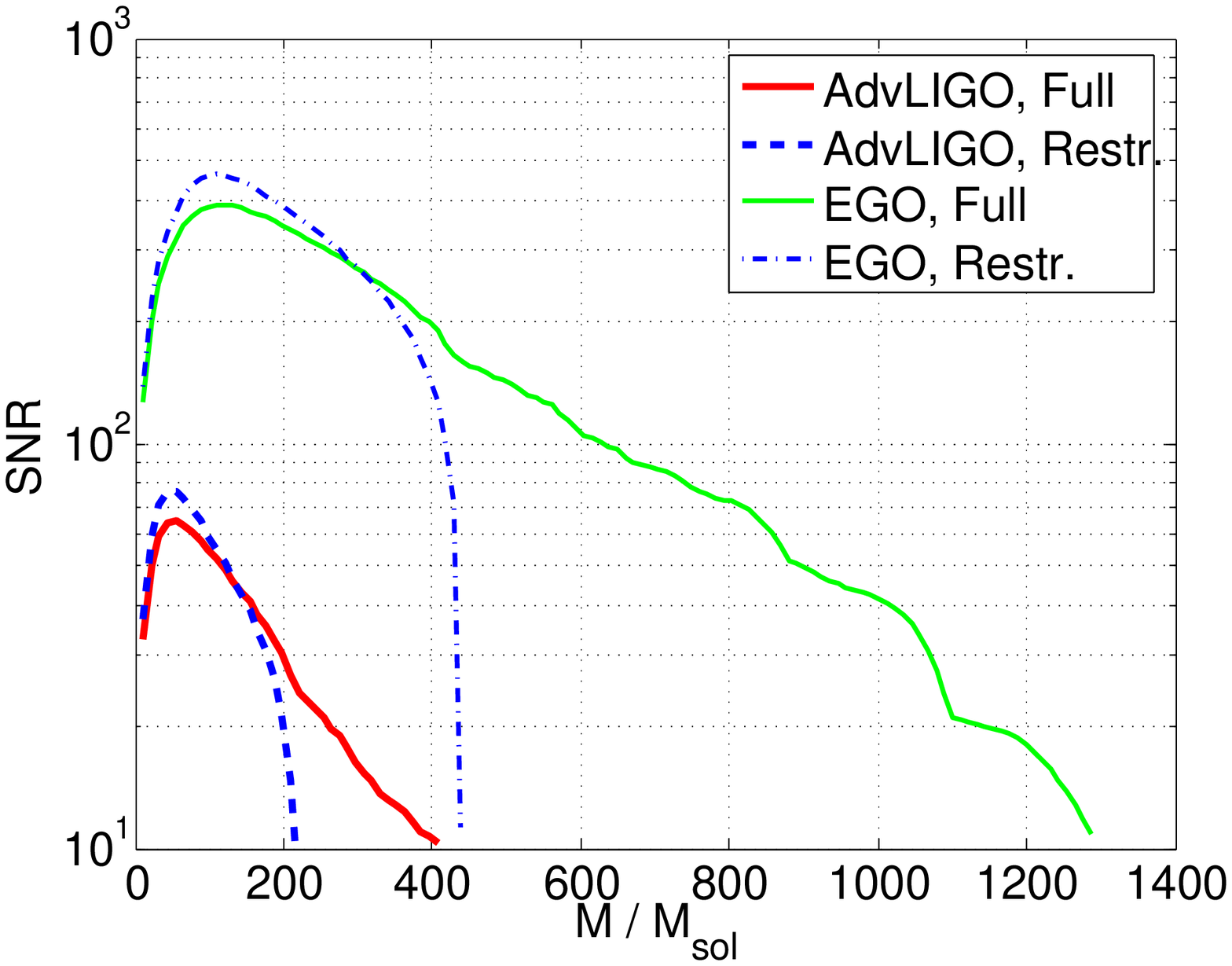}
\includegraphics[scale=0.40,angle=0]{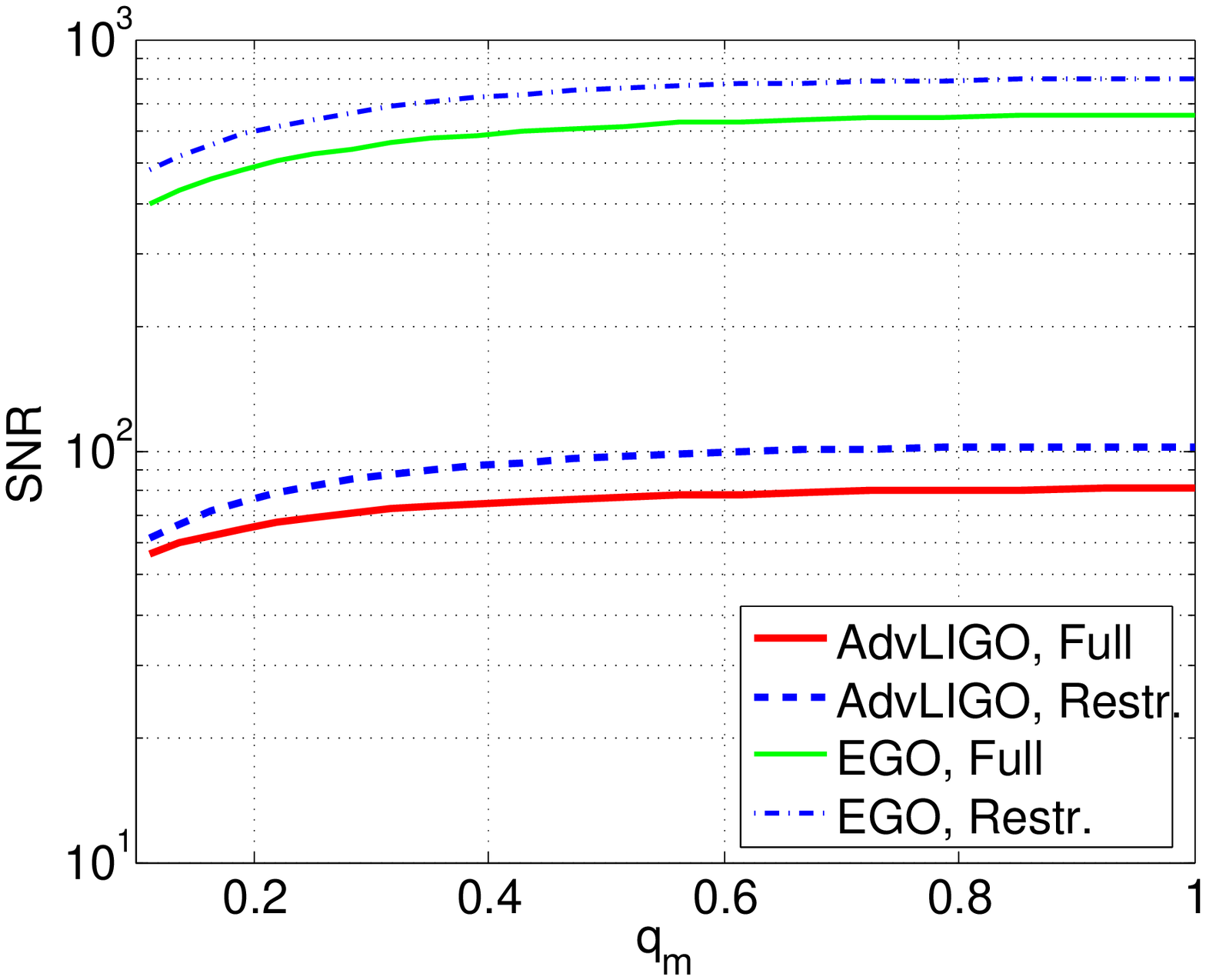}
\caption{Plots of $\rho[h]$ and $\rho[h_0]$ in Advanced LIGO and EGO as functions of total mass with fixed $q_m = m_1/m_2 = 0.1$ (left), and as functions of $q_m$ with fixed total mass $M = 100\,M_\odot$ (right). Distance is fixed at 100 Mpc and spins are set to zero. Angles were chosen arbitrarily as $\theta=\phi=\pi/6$, $\psi=\pi/4$, $\iota=\pi/3$.}
\label{f:SNR}
\end{figure}

\subsection{Signal-to-noise ratios}
\label{ss:SNRs}

The issue of signal-to-noise ratio with restricted versus amplitude-corrected waveforms has already been treated in great detail in \cite{SNRpaper} (at least for the case of zero spins) so that here we can be brief. Let us keep the mass ratio fixed at $q_m \equiv m_1/m_2=0.1$ and let $M=m_1+m_2$ vary, as in the left panel of Fig.~\ref{f:SNR}. For low masses the SNR from the restricted waveform dominates, but as mass is increased and $2 f_{LSO}$ approaches $f_s$ from above, the SNR from the full waveform catches up. For a distance of 100 Mpc, with the restricted waveforms Advanced LIGO can only detect systems with total mass up to $\sim 220\,M_\odot$ and up to double that value in EGO; heavier systems have $2 f_{LSO} < f_s$ and their signals do not enter the bandwidth. The inclusion of amplitude corrections effectively doubles the mass reach of Advanced LIGO and triples that of EGO. This is because the higher harmonics, having higher cut-off frequencies, can still penetrate the detectors' bandwidths for masses where the second harmonic does not, and they can lead to sizeable SNRs.

The right panel of Fig.~\ref{f:SNR} shows the dependence of SNR on mass ratio for a fixed total mass. Here the behavior is qualitatively the same for restricted and amplitude-corrected waveforms: In going from $q_m = 0.1$ to $q_m = 1$ the SNRs increase by a factor of a few.

\begin{figure}[htbp!]
\centering
\includegraphics[scale=0.40,angle=0]{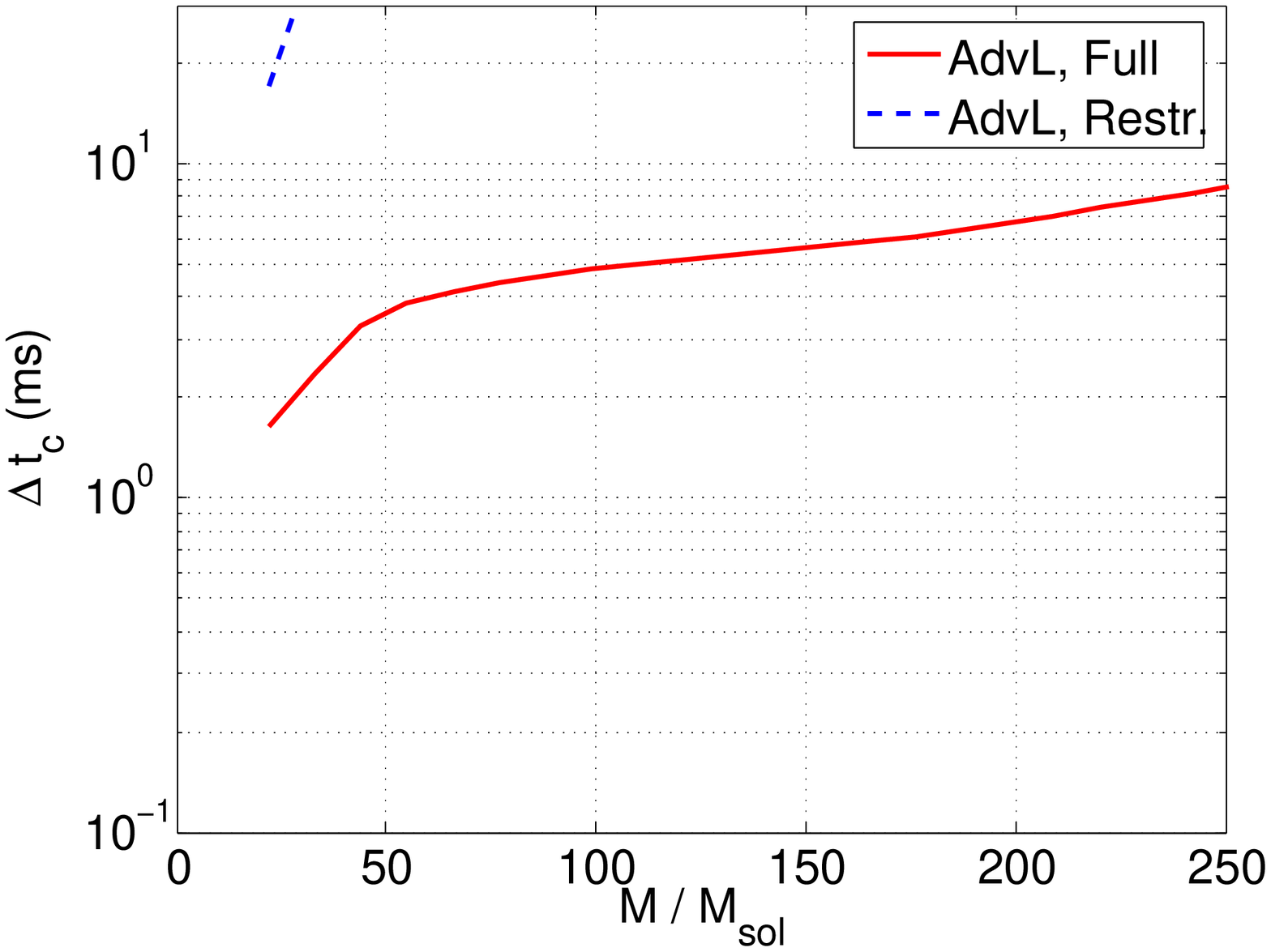}
\includegraphics[scale=0.40,angle=0]{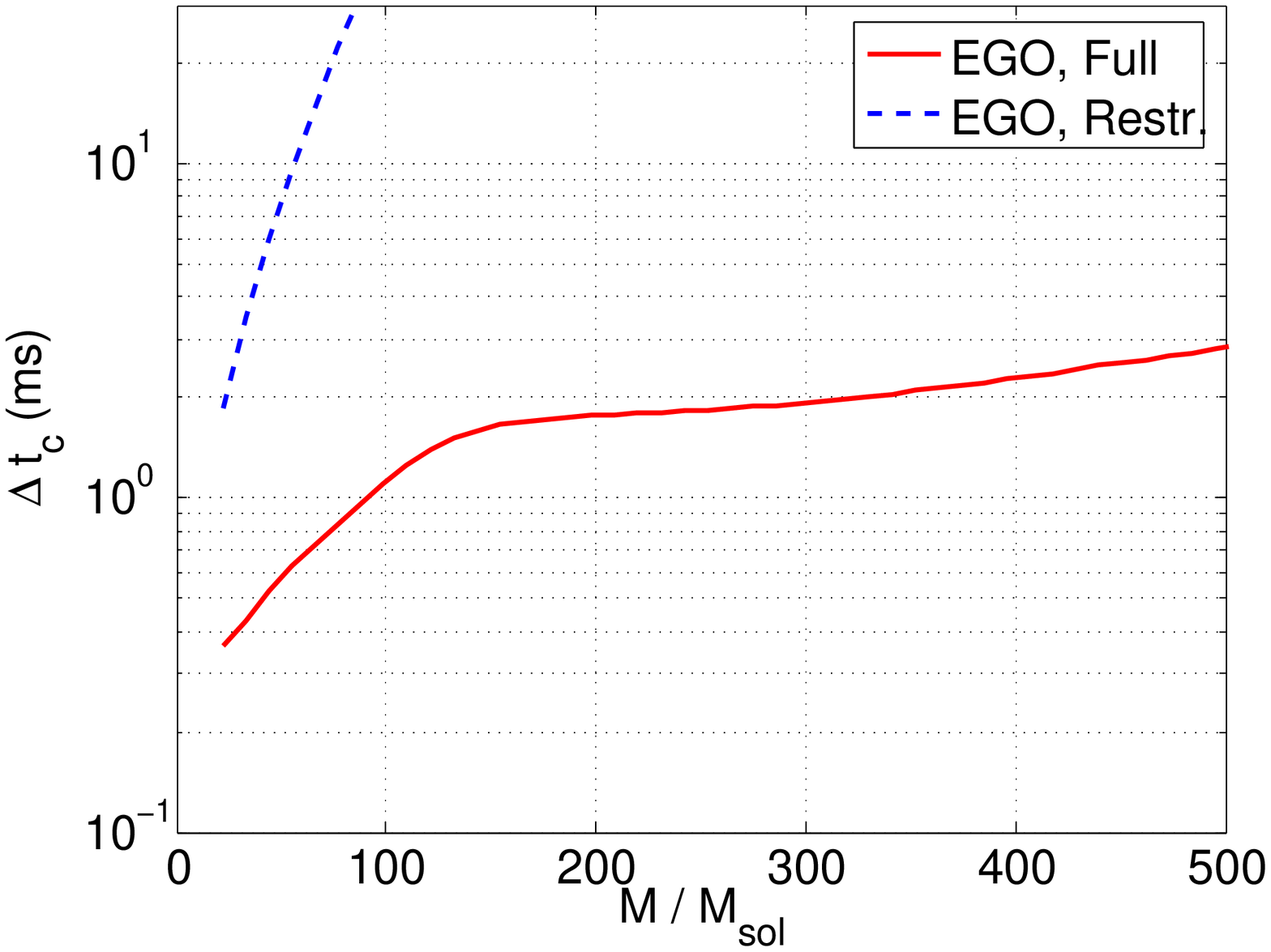}
\includegraphics[scale=0.40,angle=0]{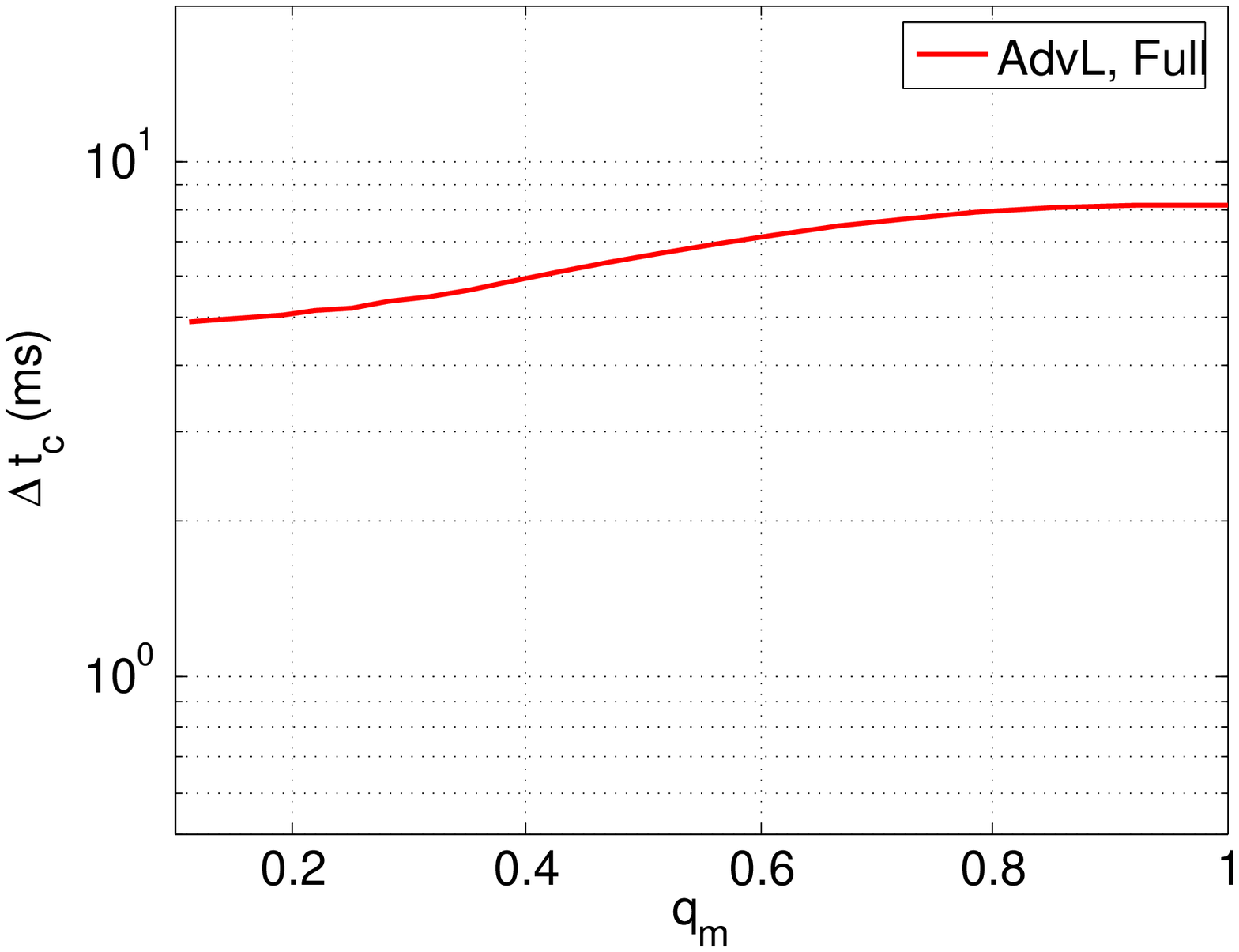}
\includegraphics[scale=0.40,angle=0]{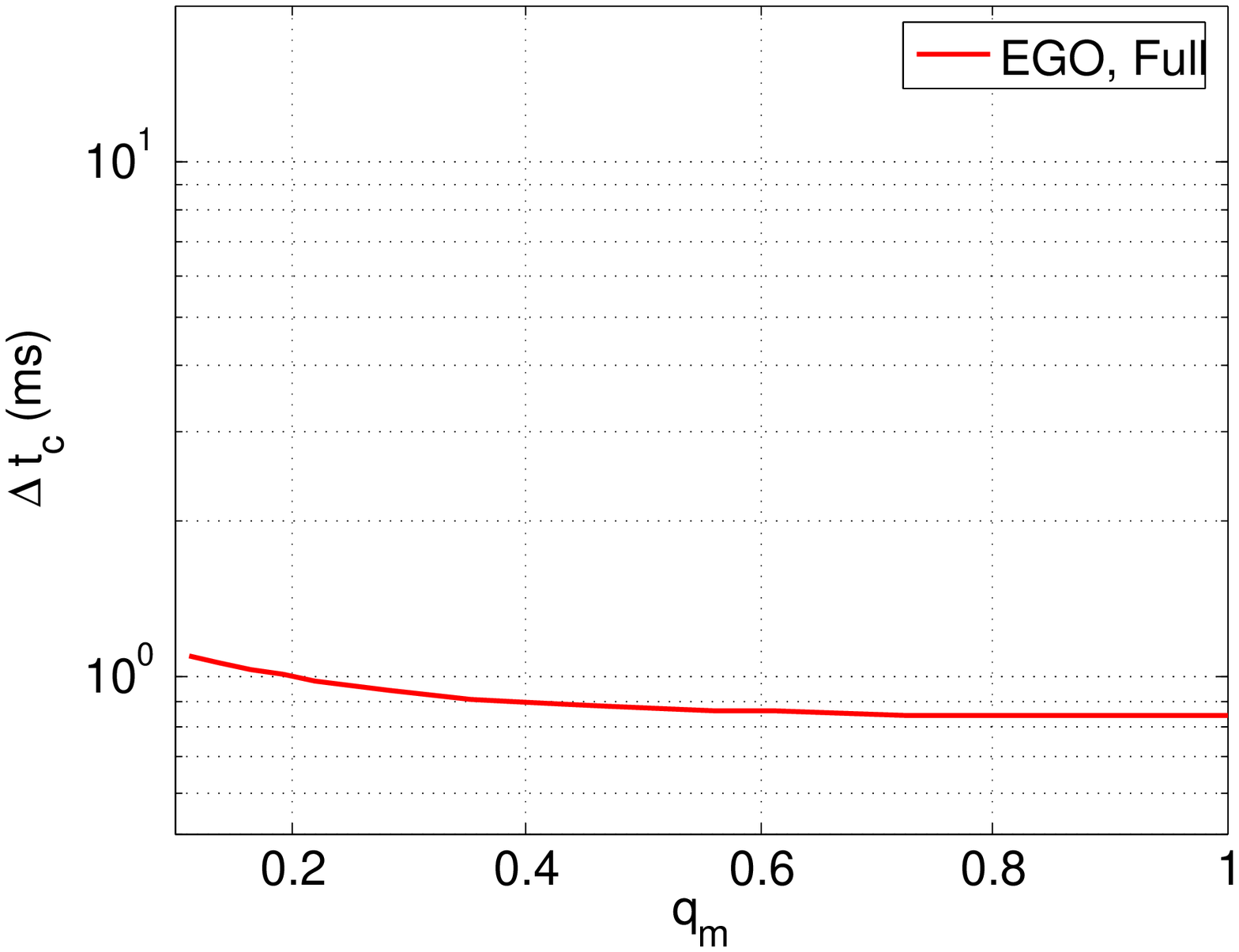}
\caption{Errors on coalescence time for varying $M$ and fixed $q_m = 0.1$ (top) and for varying $q_m$ and fixed $M = 100\,M_\odot$ (bottom), in Advanced LIGO and EGO. Here and below the distance is 100 Mpc and angles are as in the previous figure.}
\label{extrinsicfixedq}
\end{figure} 

\subsection{Extrinsic parameters}
\label{ss:extrinsic}

Let us briefly discuss the extrinsic parameters $(t_c,\ln(1/r),\psi_c,\cos(\iota),\cos(\theta),\phi,\psi)$. 

The last three of these, $(\cos(\theta),\phi,\psi)$, enter the waveform only through the beam pattern functions (\ref{beampatternfunctions}), and they do so in very similar ways. As a result, these three variables have large correlations (in absolute value) with each other and very small correlations with all of the other variables. Indeed, using the Fisher matrix based on all of the parameters in (\ref{coordinates}), typically one has
\be
c^{\cos(\theta), \phi} = -c^{\cos(\theta), \psi} = -c^{\phi, \psi} = 1,
\label{correlations}
\ee
with an accuracy of eight significant digits,
and the correlations of these parameters with the other eight tend to stay below $10^{-3}$ in absolute value. These angles are unmeasurable in practice, with $\Delta\cos(\theta) \gg 1$, $\Delta\phi \gg 2\pi$, and $\Delta\psi \gg 2\pi$. The distance $r$ is also effectively unmeasurable, $\Delta r/r \gg 1$, as is the phase at coalescence $\psi_c$. As mentioned before, $\Delta\cos(\iota)$ tends to be reasonably small ($\sim 0.2$ in Advanced LIGO and $\sim 0.02$ in EGO), but the information it carries is not of much interest by itself.

This leaves the coalescence time $t_c$. As can be seen in Fig.~\ref{extrinsicfixedq}, for small ($\lesssim 50\,M_\odot$) masses the restricted waveforms give errors of a few tens of milliseconds in Advanced LIGO and a few milliseconds in EGO; as total mass is increased, $t_c$ quickly becomes unmeasurable in practice. By contrast, using the full $(2.5,2.5)$PN waveforms, the errors stay below 10 ms for $M \leq 250\,M_\odot$ in Advanced LIGO and below 3 ms for $M \leq 500\,M_\odot$ in EGO.\footnote{At first glance our numerical results for the restricted waveform in both this and the following subsections may seem at odds with \cite{Arunetal,PoissonWill}. However, (i) in \cite{Arunetal} no spin-related parameters were taken into account, which, as shown in \cite{PoissonWill}, leads to smaller errors; and (ii) in \cite{PoissonWill} a somewhat idealized noise power spectral density was used which is quite different from the more accurate analytic PSDs now available. Leaving out $\beta$ and $\sigma$ we were able to reproduce the results of \cite{Arunetal} for both Initial and Advanced LIGO, and those of \cite{PoissonWill} were retrieved when using their PSD.} For a fixed total mass of $M = 100\,M_\odot$ and increasing $q_m$, there is little change in $\Delta t_c$.

The overall situation remains qualitatively the same irrespective of mass ratio. Thus, except for the inclination angle, the parameters that give the binary's position and orientation with respect to the detector are unmeasurable in practice, both with Advanced LIGO and EGO. (Things would of course be different with a network of detectors \cite{Tinto}, but that is outside the scope of this paper.) With amplitude-corrected waveforms the time of coalescence is reasonably well determined over a large range of masses, while with the restricted ones $t_c$ can be measured for stellar mass binaries only.

\begin{figure}[htbp!]
\centering
\includegraphics[scale=0.40,angle=0]{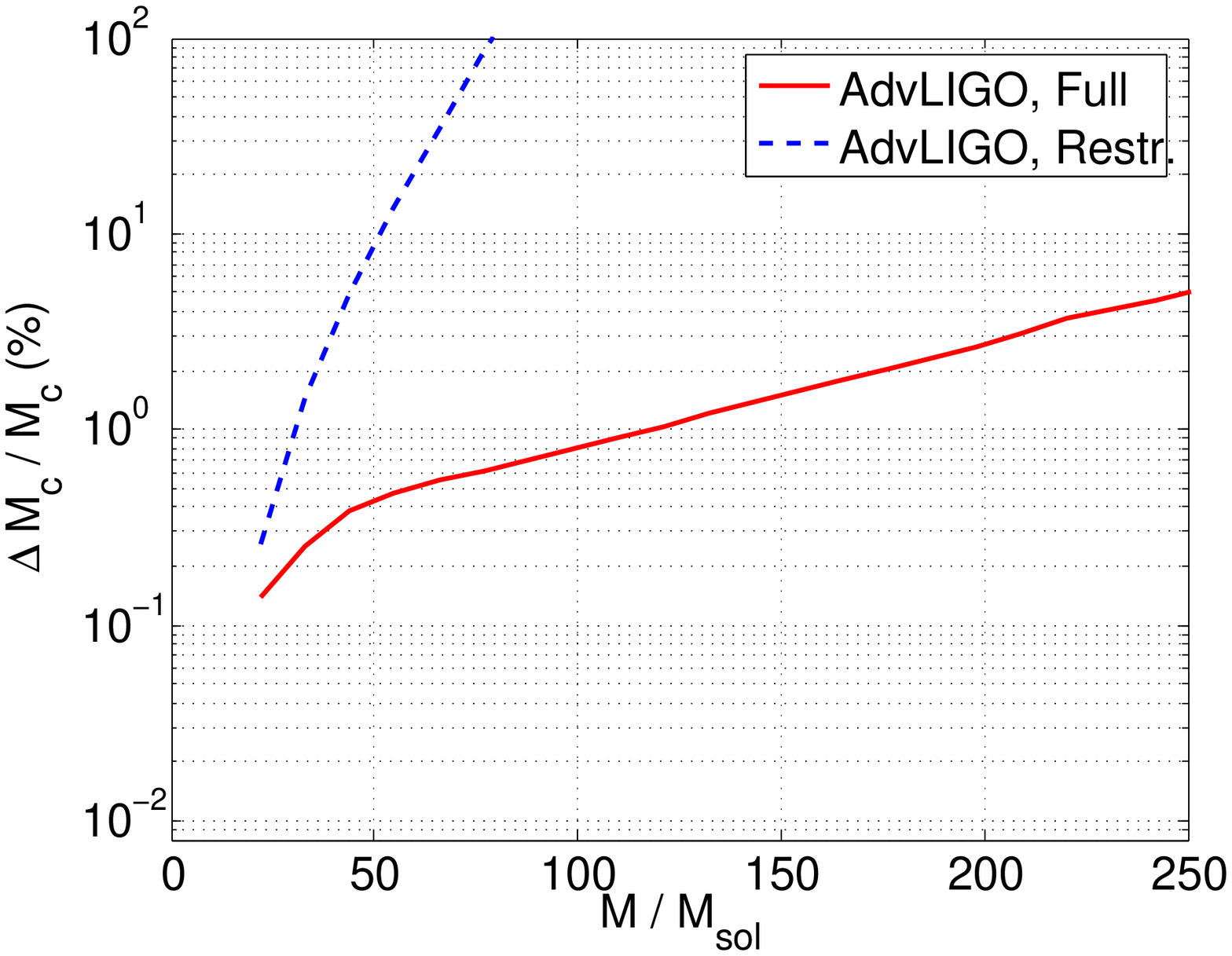}
\includegraphics[scale=0.40,angle=0]{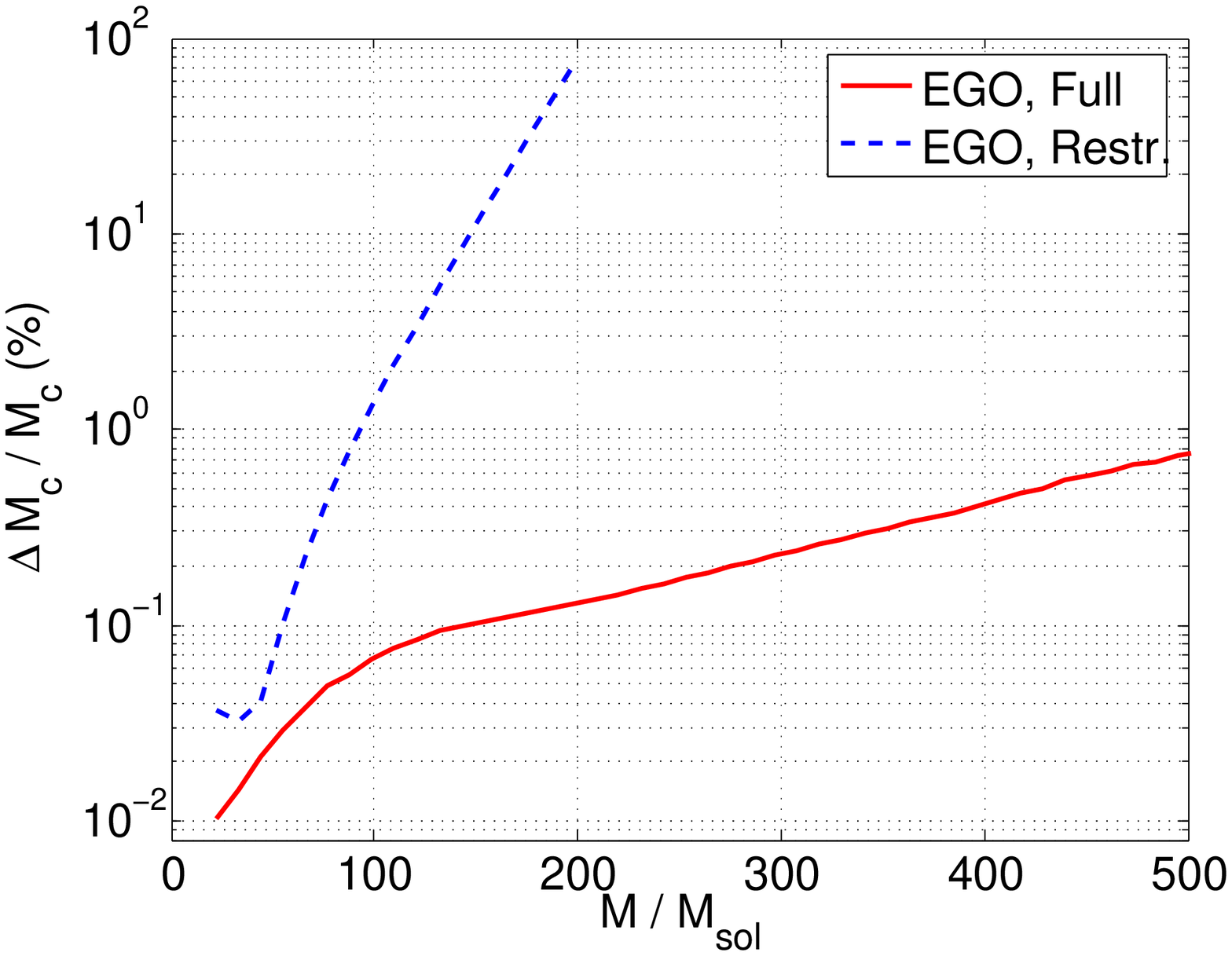}
\includegraphics[scale=0.40,angle=0]{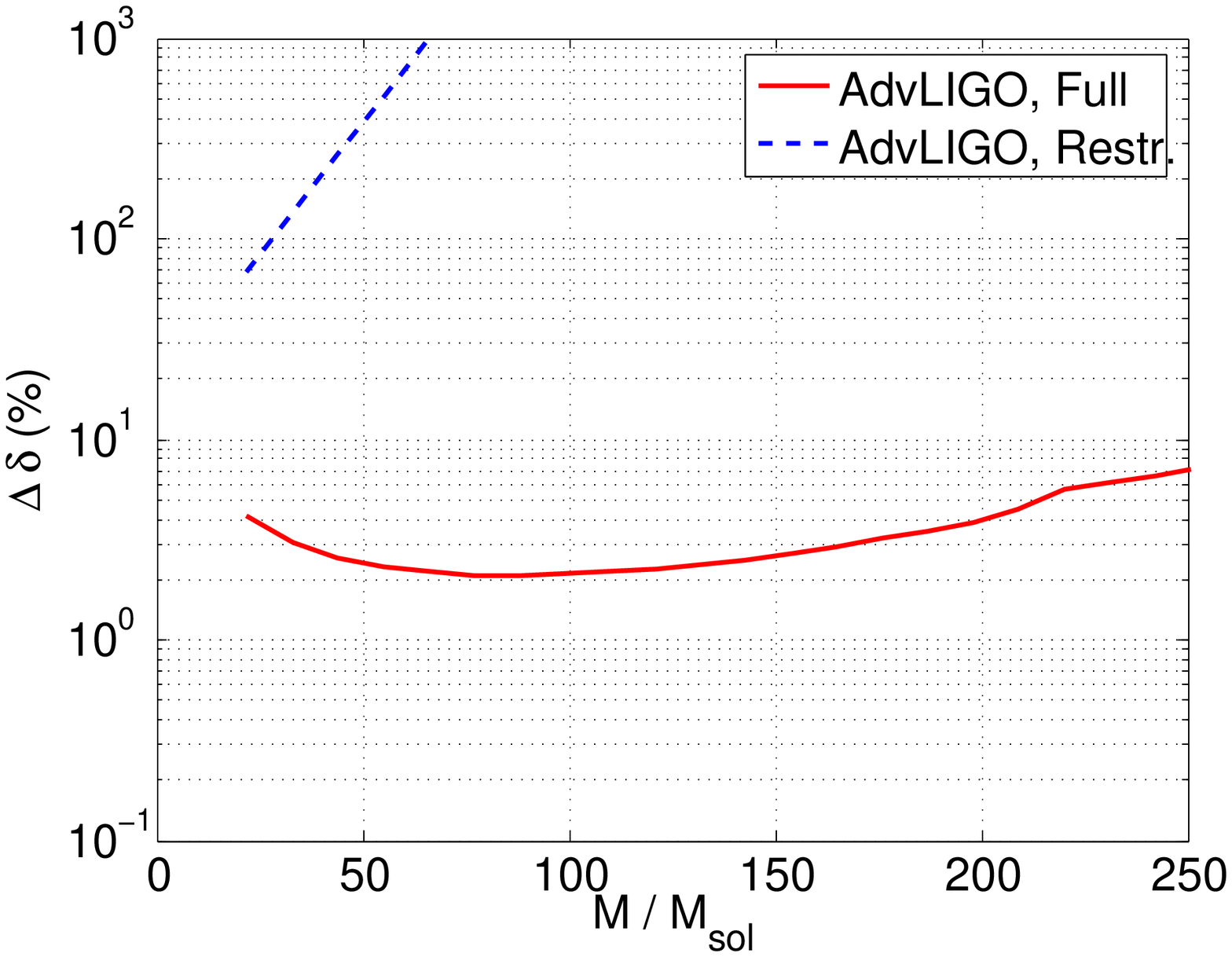}
\includegraphics[scale=0.40,angle=0]{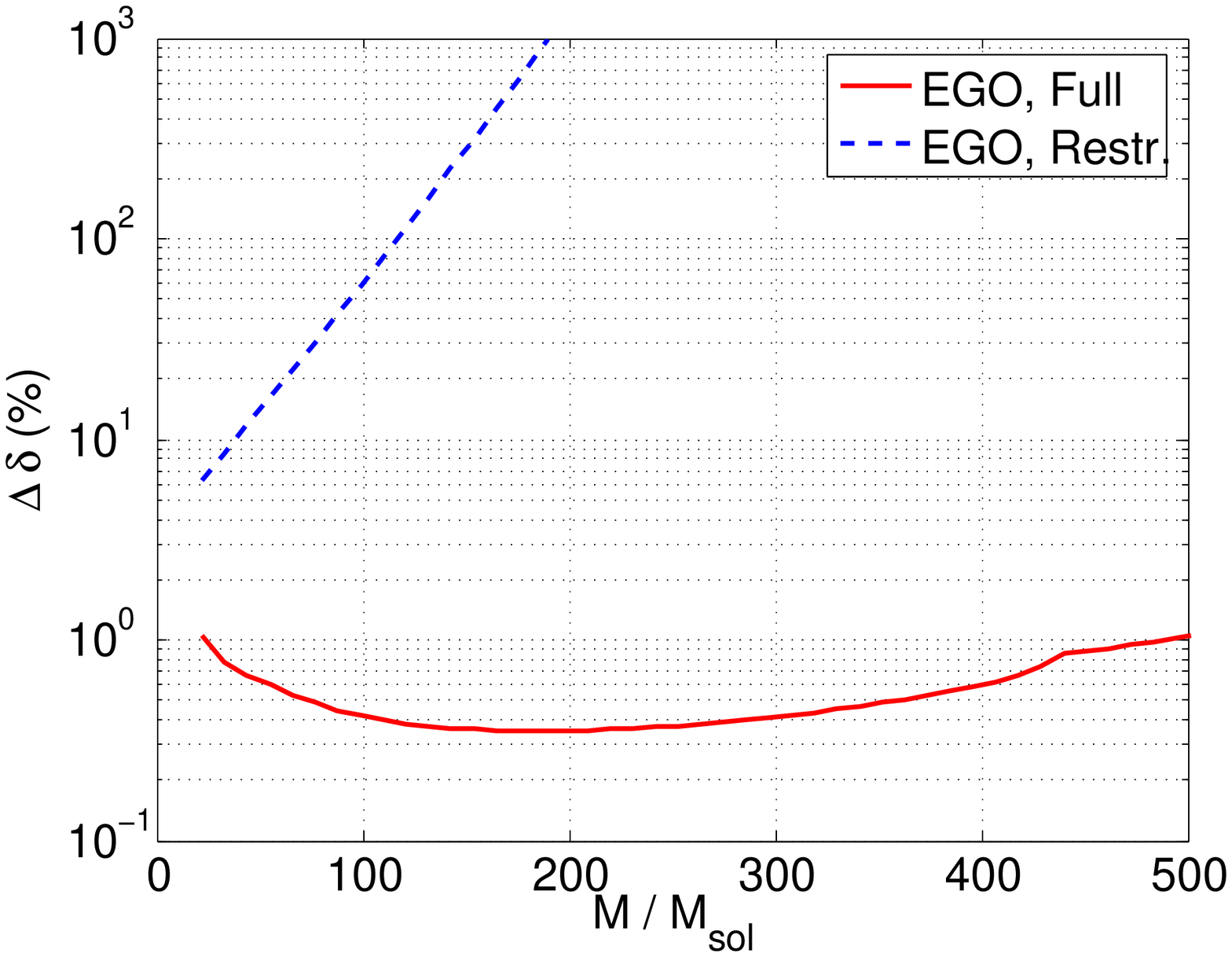}
\includegraphics[scale=0.40,angle=0]{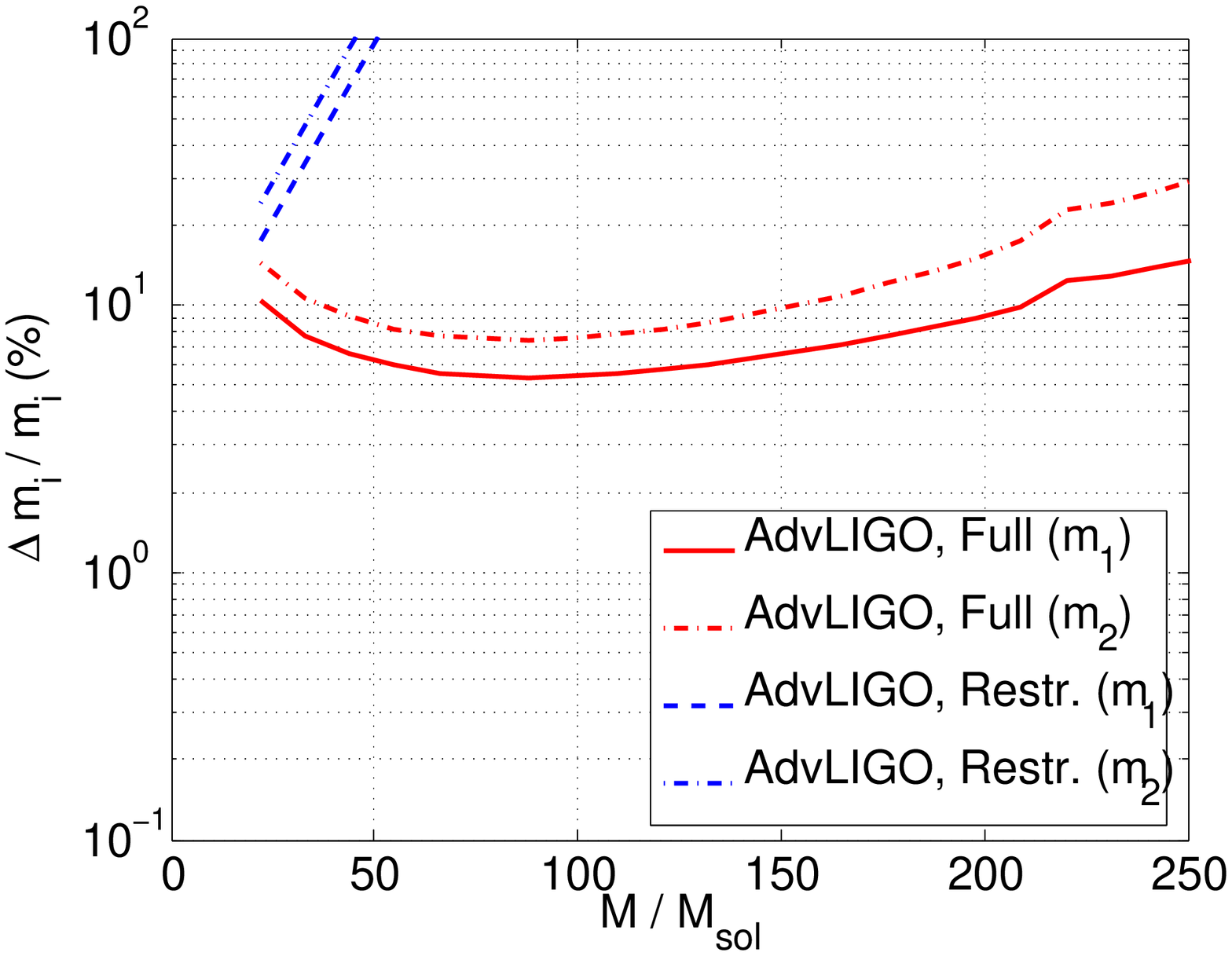}
\includegraphics[scale=0.40,angle=0]{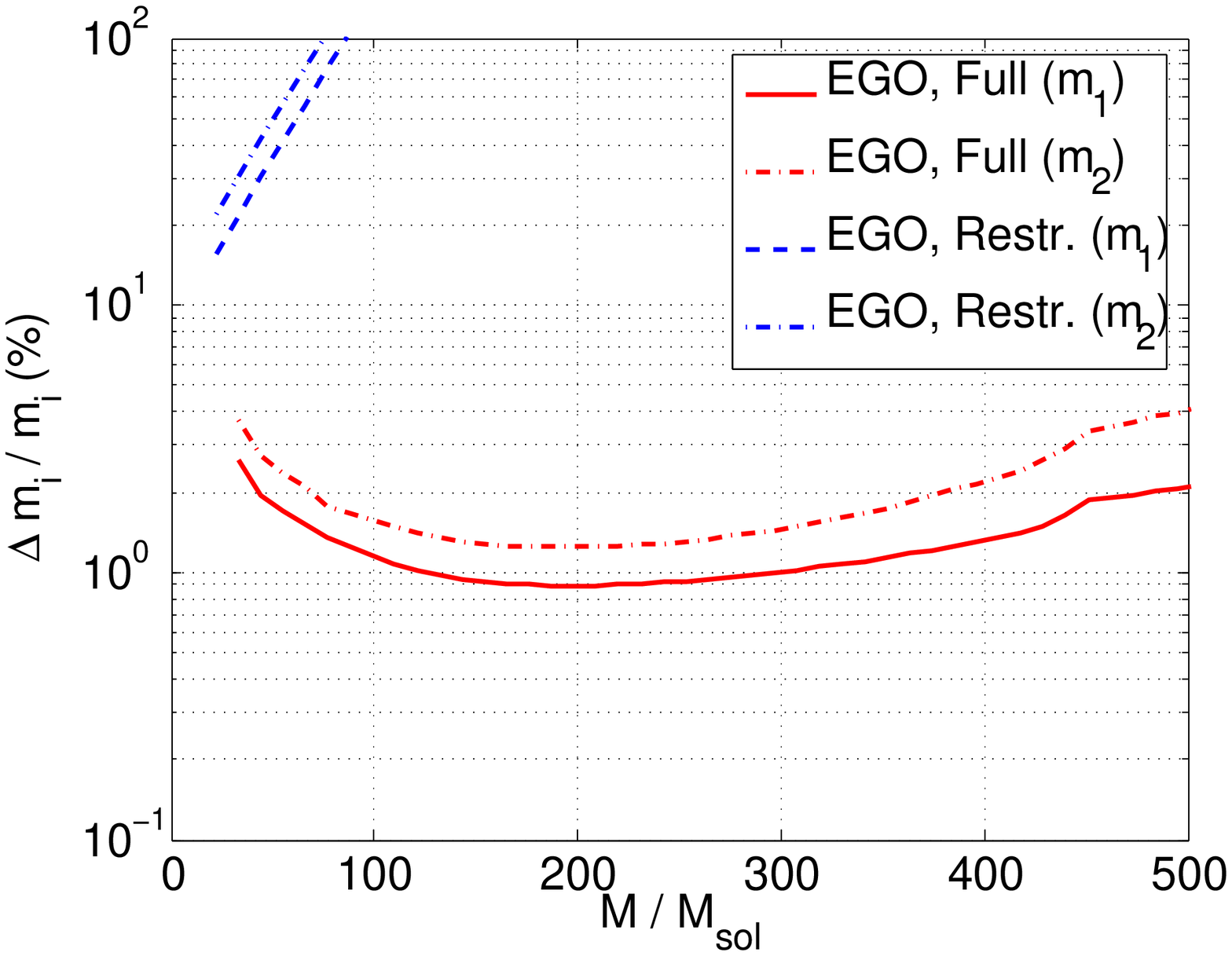}
\caption{Errors on chirp mass, relative mass difference, and component masses for $q_m=0.1$ in Advanced LIGO and EGO.}
\label{compmassesfixedq}
\end{figure}  

\begin{figure}[htbp!]
\centering
\includegraphics[scale=0.40,angle=0]{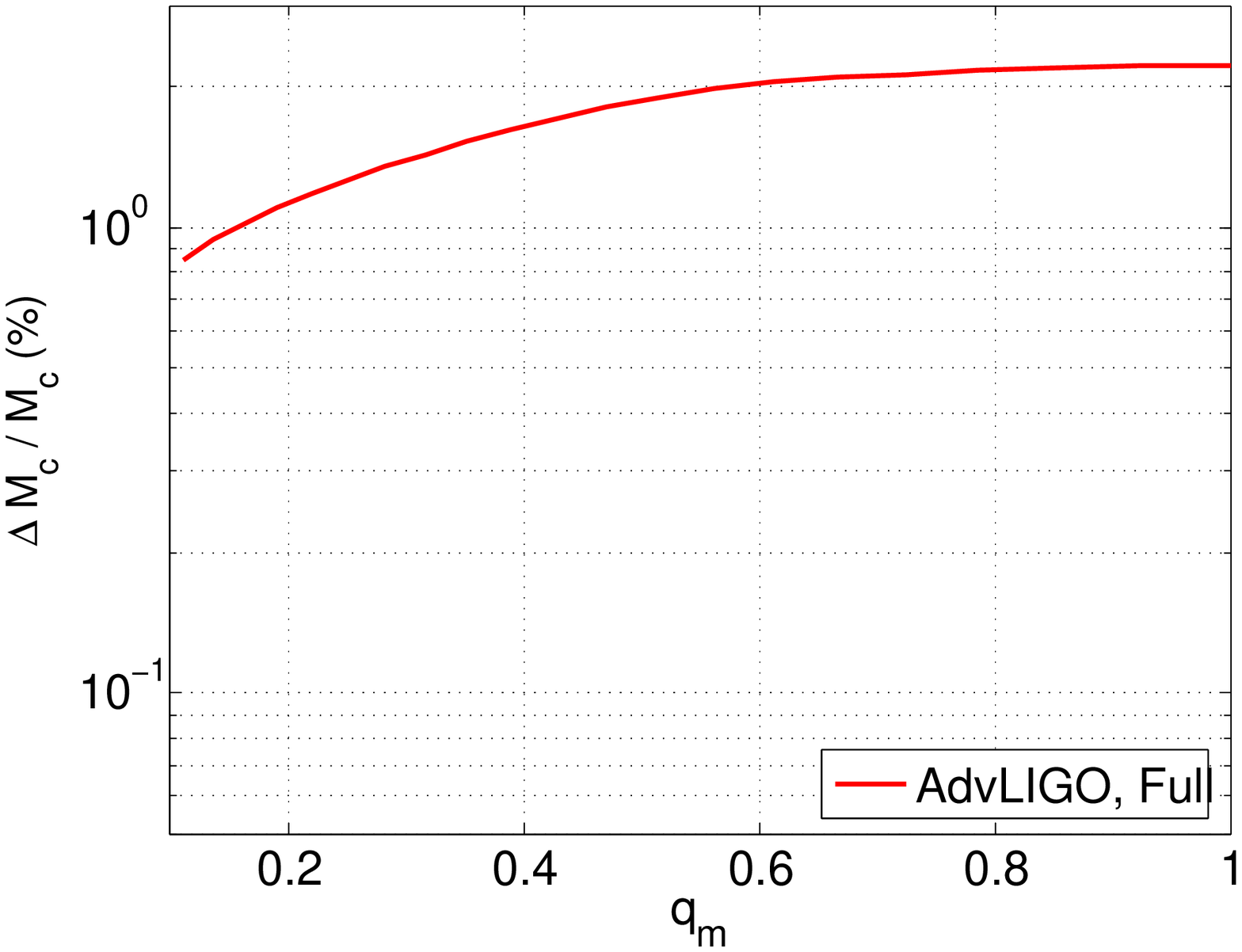}
\includegraphics[scale=0.40,angle=0]{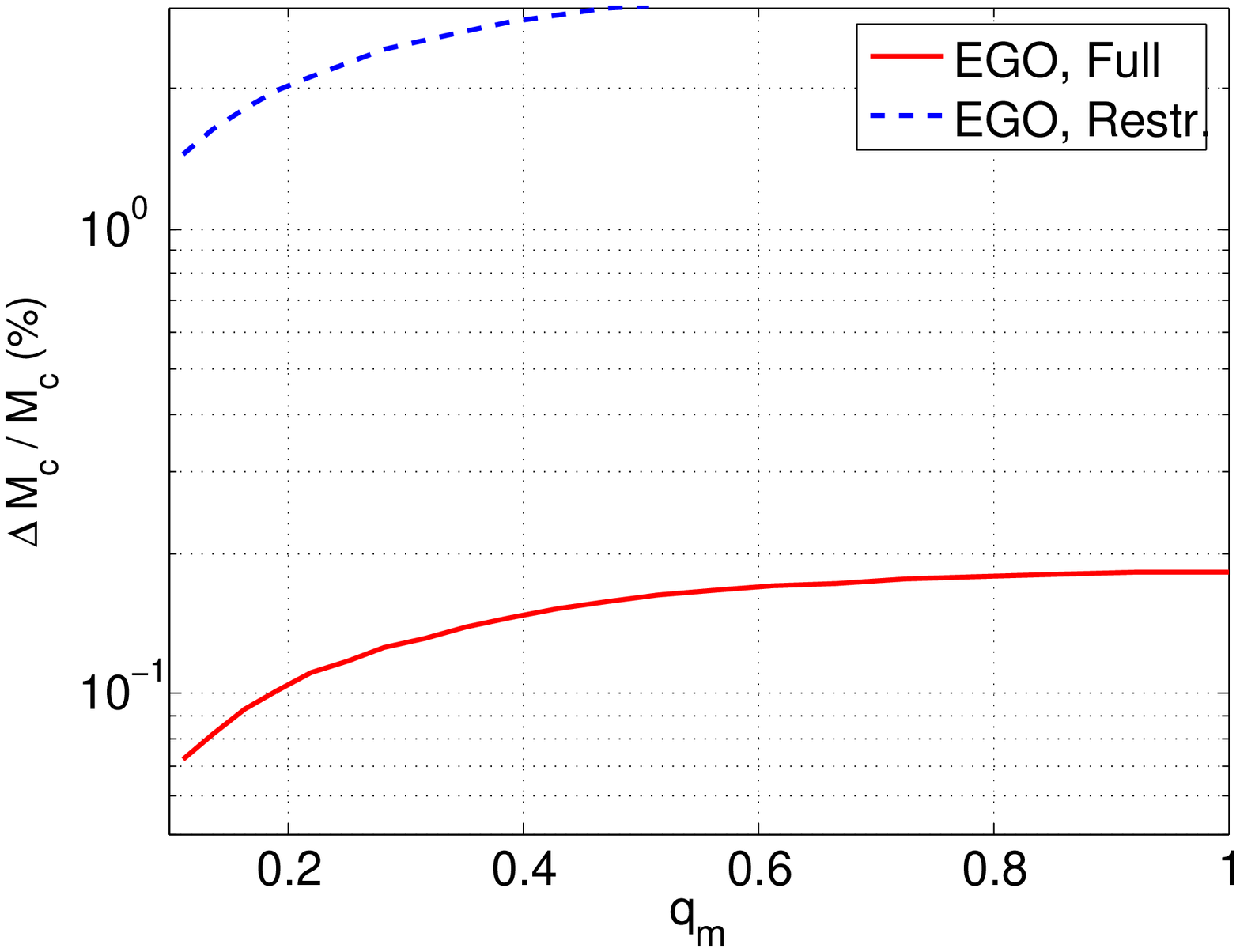}
\includegraphics[scale=0.40,angle=0]{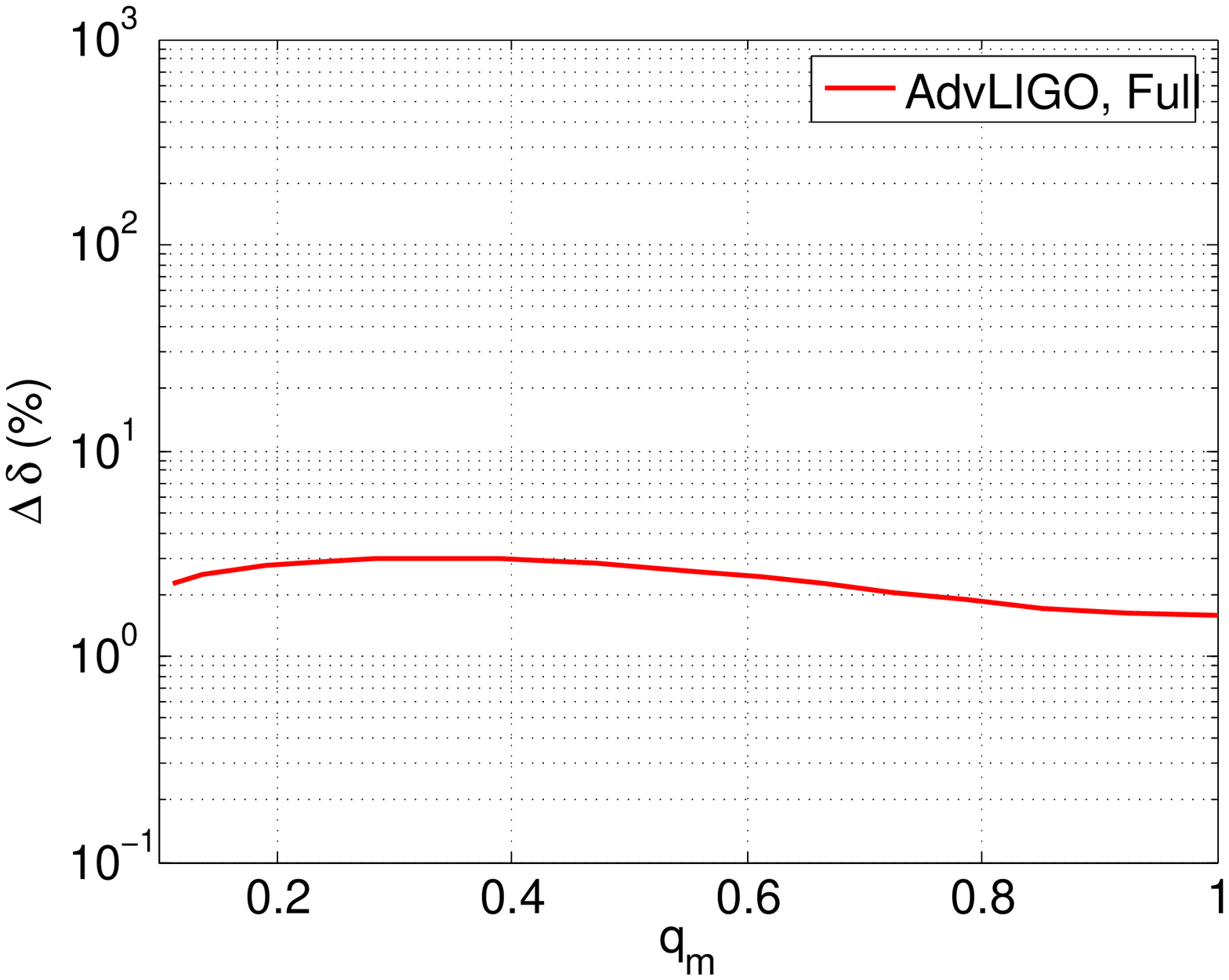}
\includegraphics[scale=0.40,angle=0]{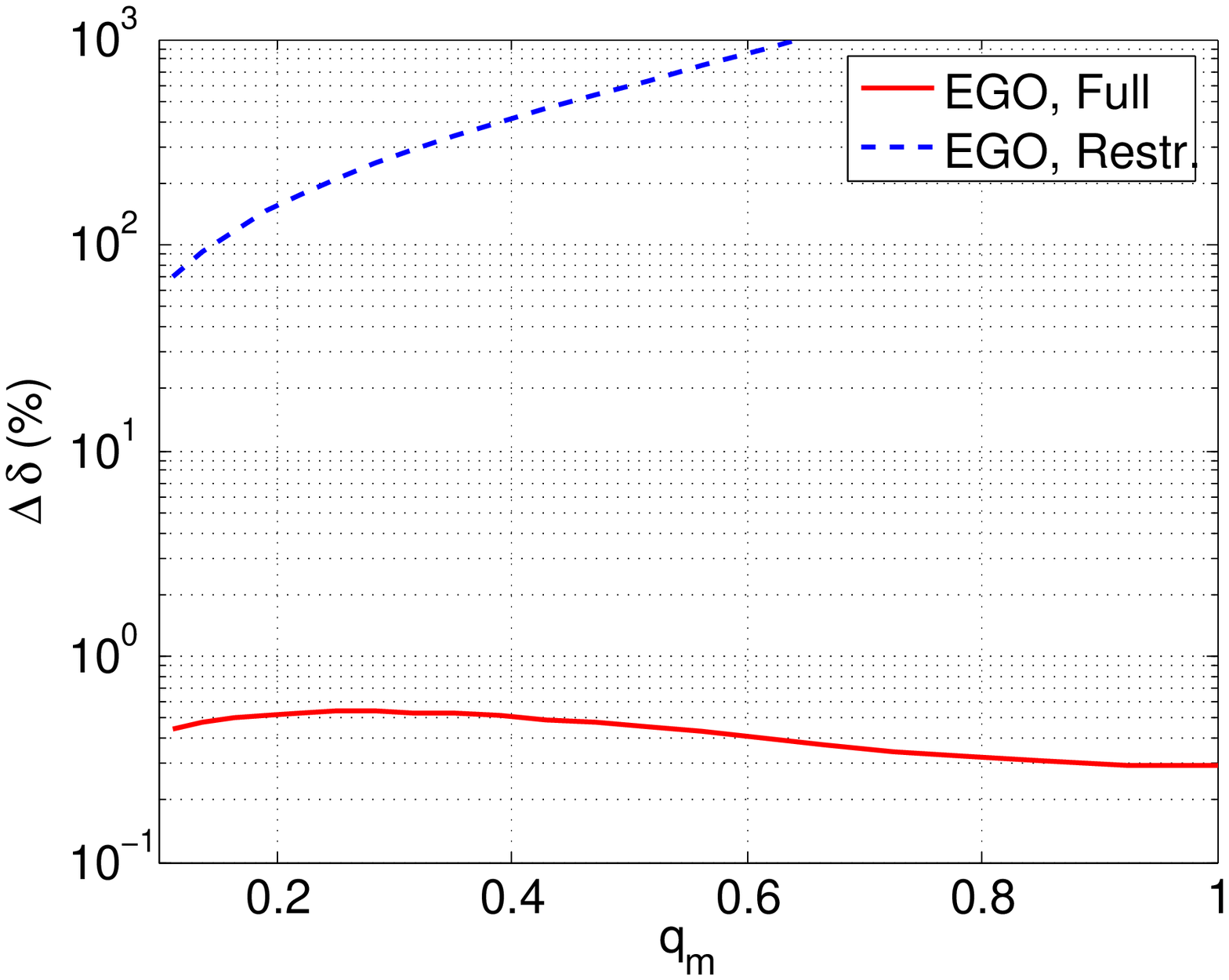}
\includegraphics[scale=0.40,angle=0]{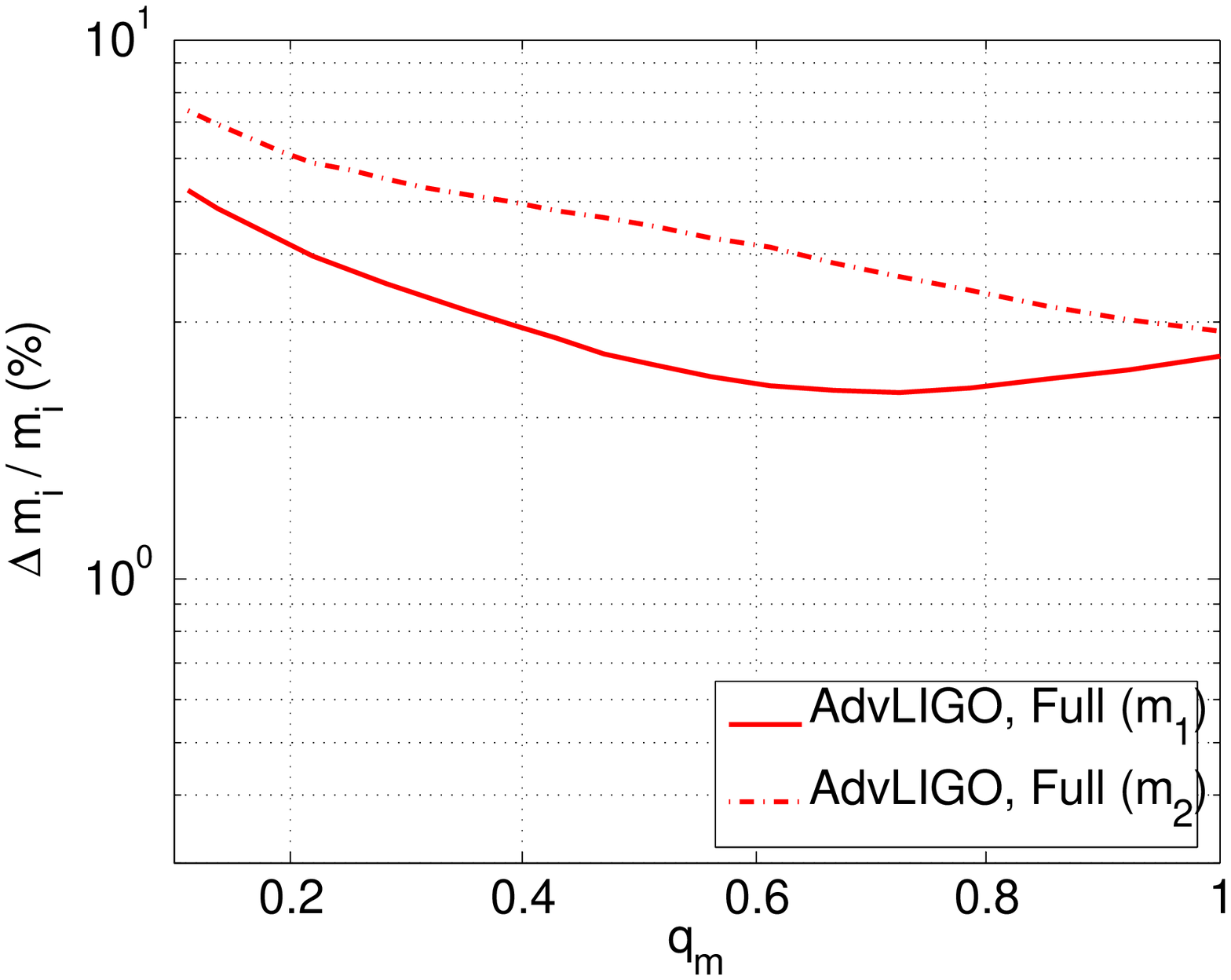}
\includegraphics[scale=0.40,angle=0]{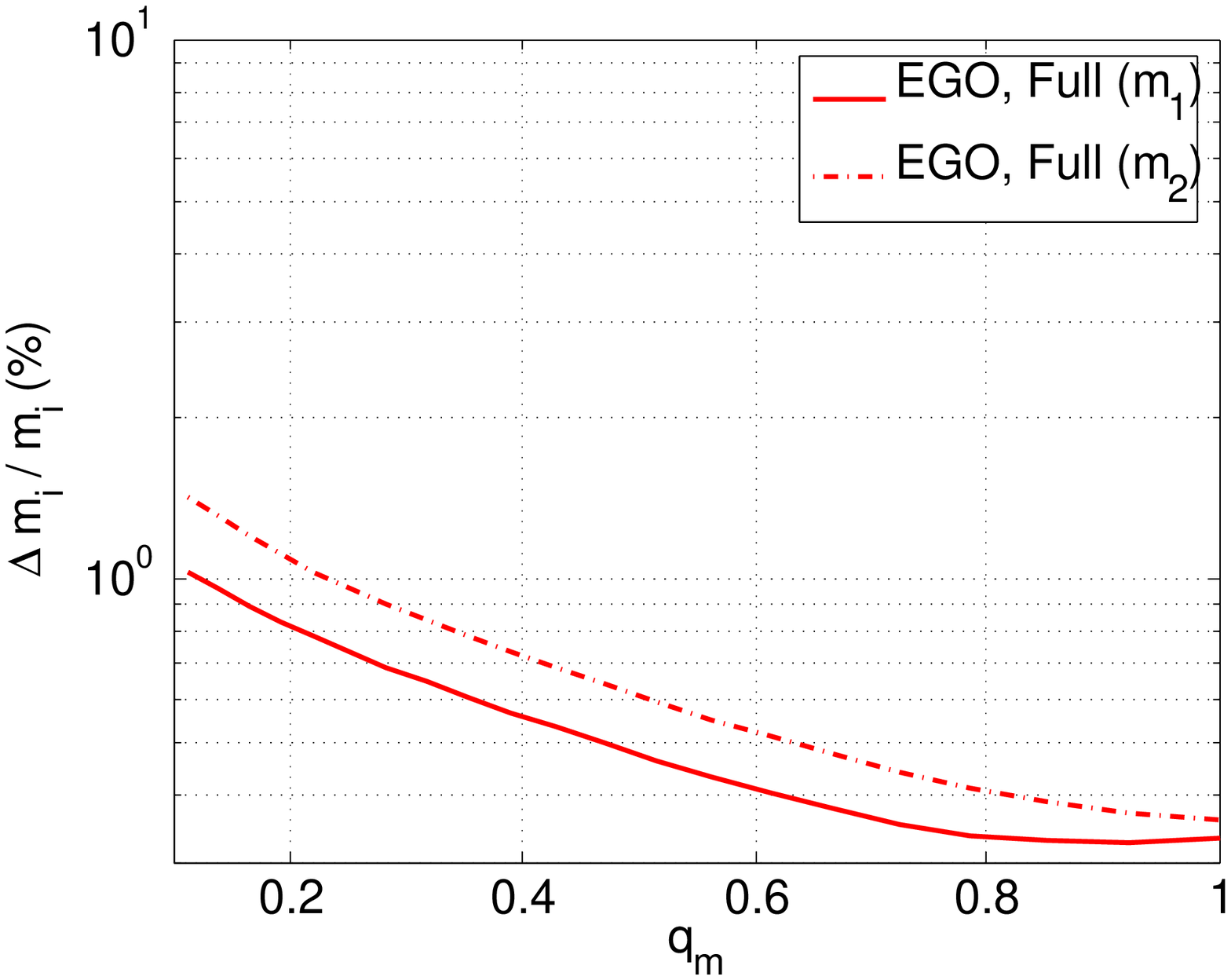}
\caption{Errors on chirp mass, relative mass difference, and component masses for $M=100\,M_\odot$ in Advanced LIGO and EGO.}
\label{compmassesfixedM}
\end{figure}

\subsection{Intrinsic parameters} 
\label{ss:intrinsic}

\begin{figure}[htbp!]
\centering
\includegraphics[scale=0.40,angle=0]{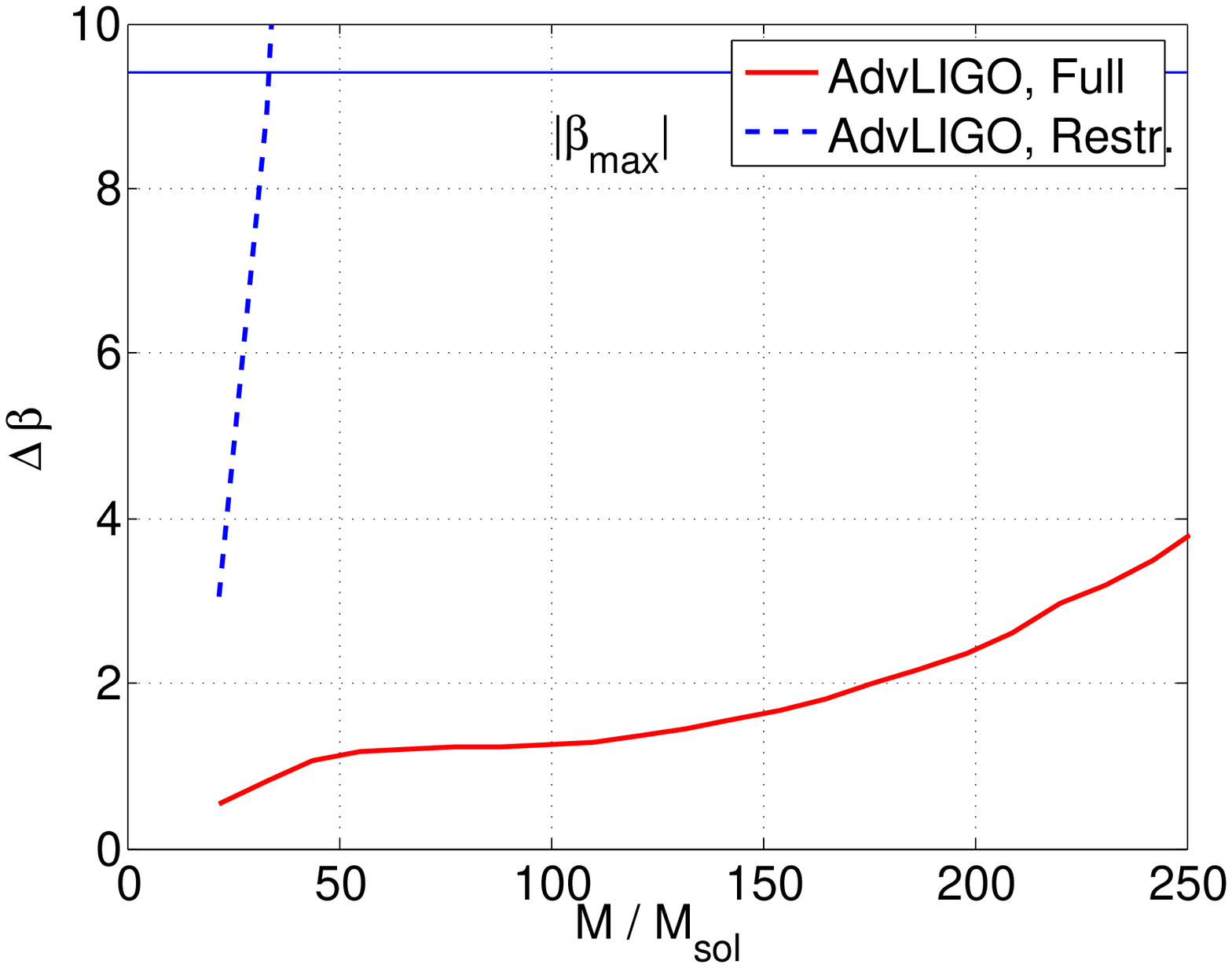}
\includegraphics[scale=0.40,angle=0]{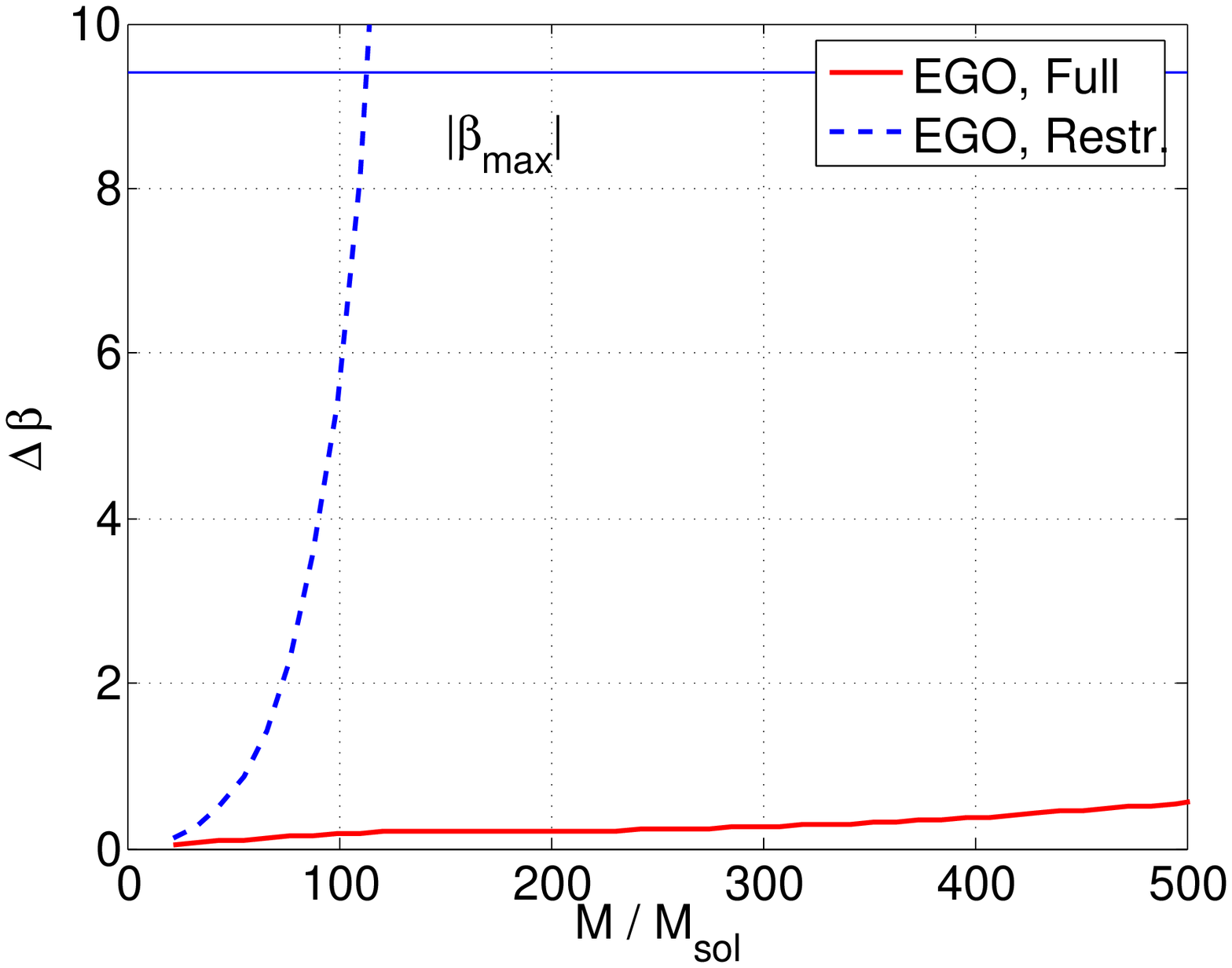}
\includegraphics[scale=0.40,angle=0]{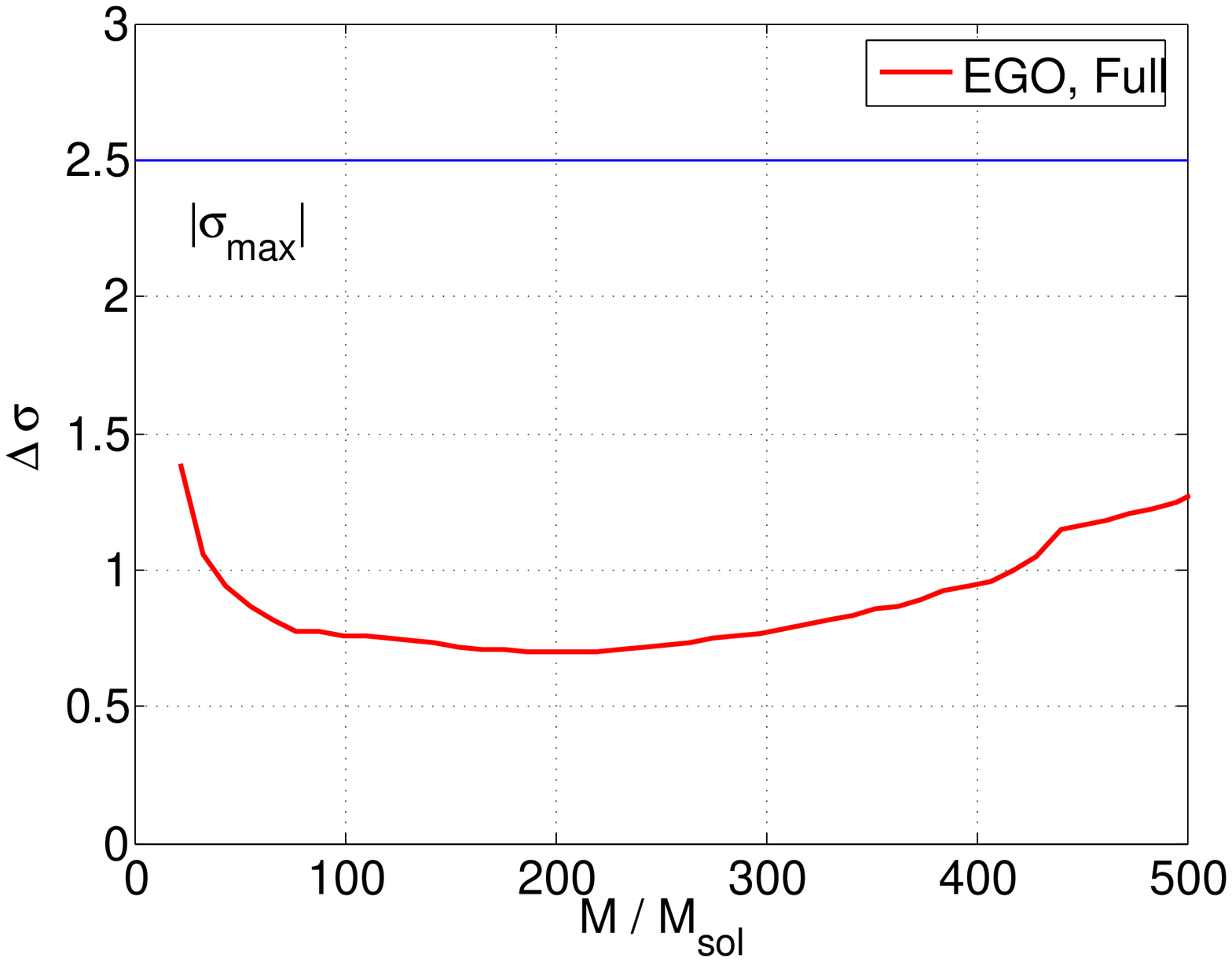}
\caption{Errors on the spin-orbit parameter $\beta$ and the spin-spin parameter $\sigma$ with increasing total mass, keeping the mass ratio fixed at $q_m=0.1$. The largest values $\beta$ and $\sigma$ can take for sub-extremal black holes are shown by the horizontal lines.}
\label{spinsFigure}
\end{figure}  

\begin{figure}[htbp!]
\centering
\includegraphics[scale=0.40,angle=0]{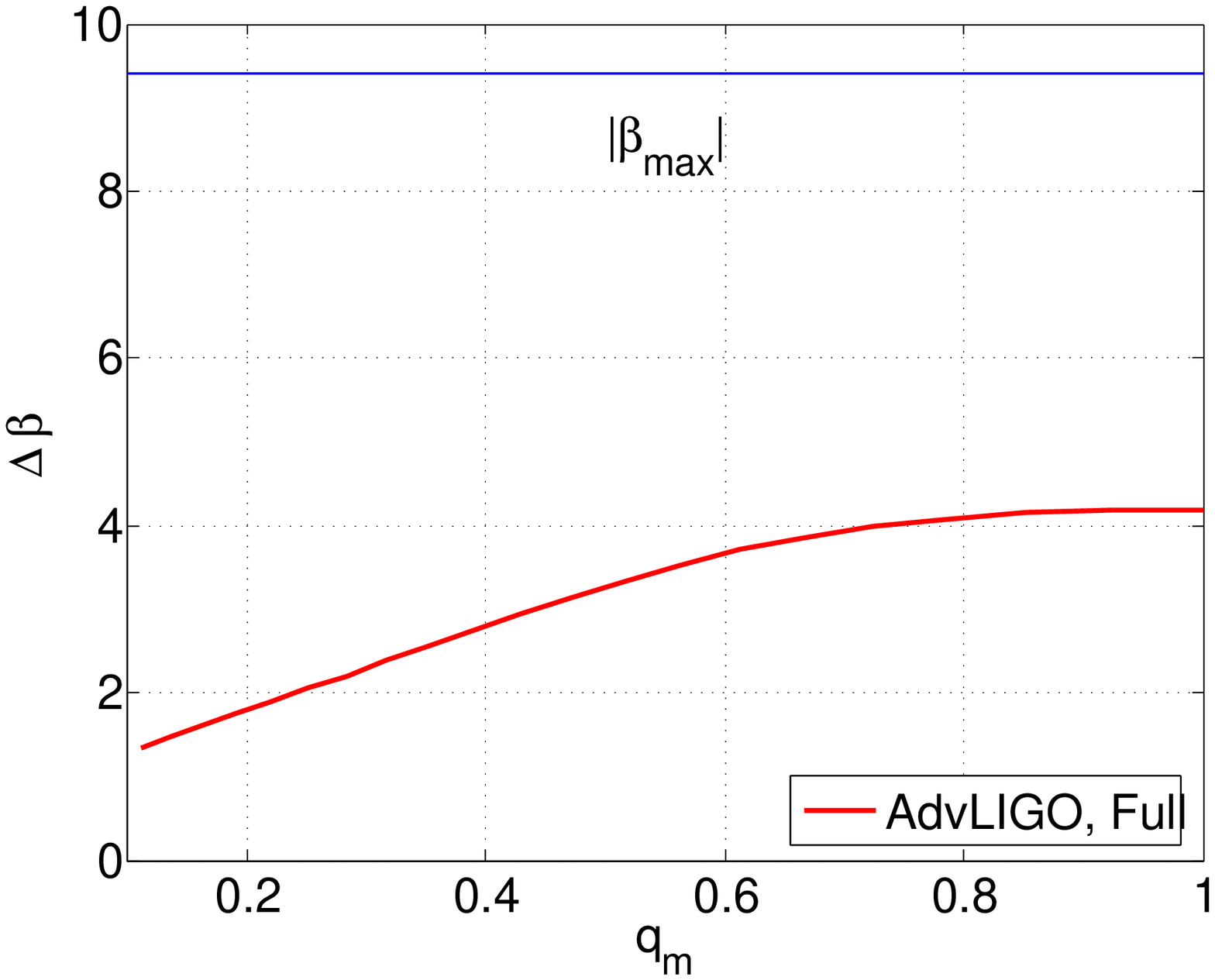}
\includegraphics[scale=0.40,angle=0]{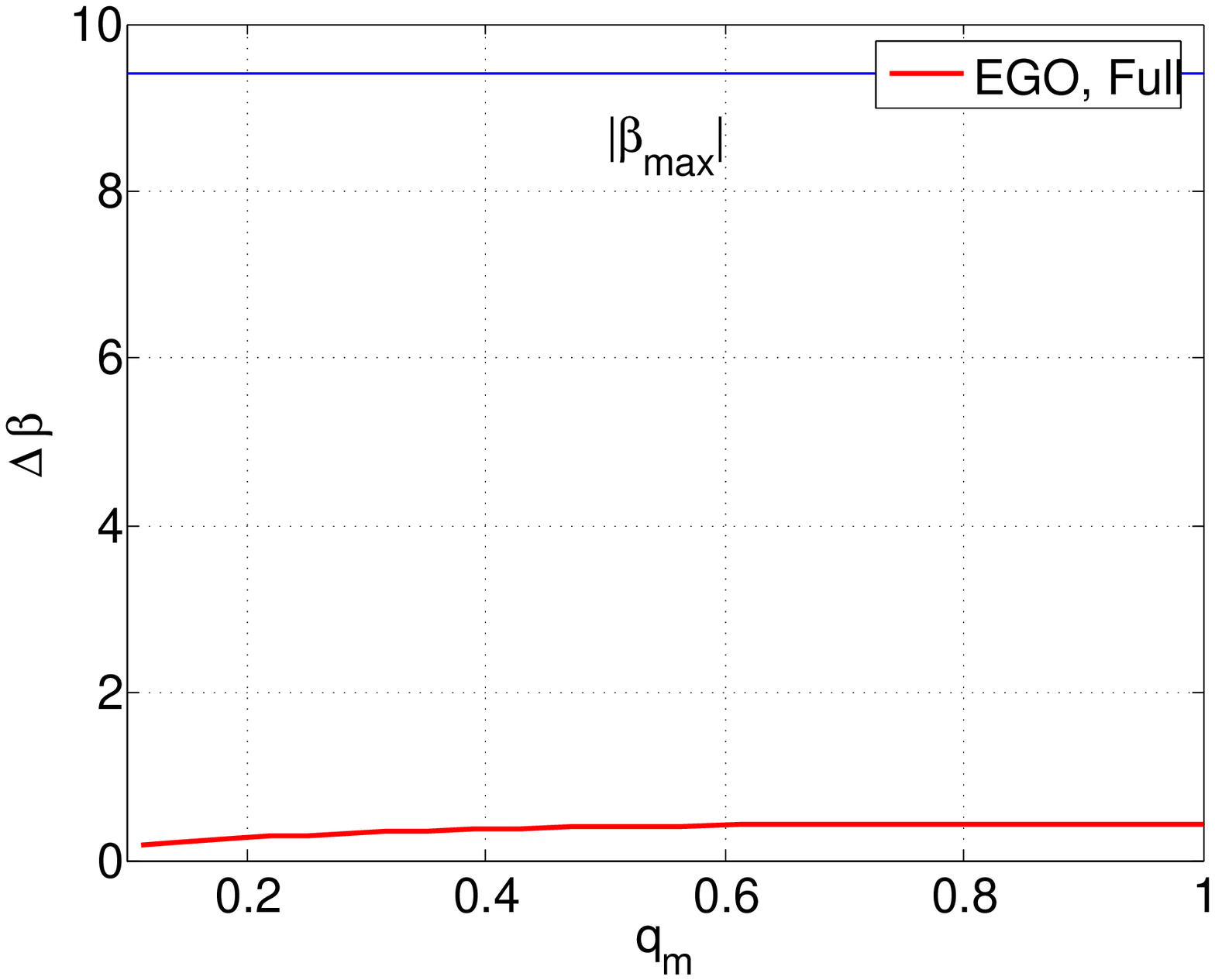}
\includegraphics[scale=0.40,angle=0]{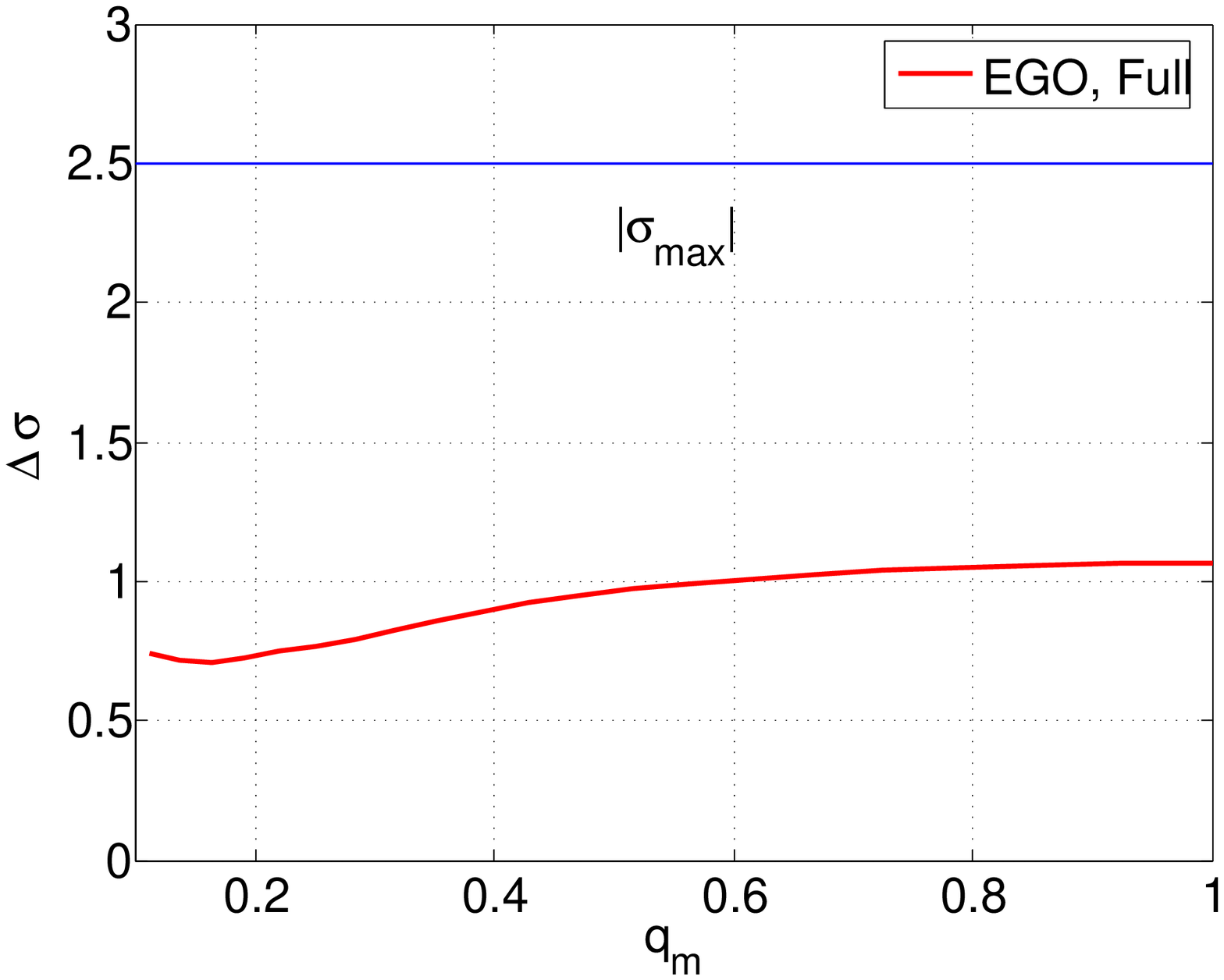}
\caption{Errors on the spin-orbit parameter $\beta$ and the spin-spin parameter $\sigma$ with increasing mass ratio, keeping total mass fixed at $M=100\,M_\odot$. Again the largest possible values of $\beta$ and $\sigma$ for sub-extremal black holes are shown by the horizontal lines.}
\label{spinsFigureq}
\end{figure}  

We now turn to the errors on the more interesting intrinsic parameters $(\ln(\M),\delta,\beta,\sigma)$. 

For the case of fixed $q_m=0.1$ and varying $M$, the errors on $\M$ are shown in Fig.~\ref{compmassesfixedq} using the restricted and $(2.5,2.5)$PN amplitude-corrected waveforms in Advanced LIGO and EGO. As with $t_c$, using restricted waveforms the errors tend to be small for low total mass but rise very steeply with increasing $M$, becoming effectively unmeasurable as one moves beyond the mass range of $[2\,M_\odot,100\,M_\odot]$ corresponding to stellar mass binaries. Using the amplitude-corrected waveforms they stay below 5\% up to $M = 250\,M_\odot$ in Advanced LIGO and below 1\% up to $M = 500\,M_\odot$ in EGO. The corresponding numbers for $\Delta\delta$ happen to be quite similar when expressed in percent.\footnote{Although $\delta$ is dimensionless, we opted to express its errors in percent. They tend to be small numbers, which otherwise would have led to inconvenience when presenting them in tables, as we will do later on.} Fig.~\ref{compmassesfixedM} shows how errors change with varying $q_m$ for $M=100\,M_\odot$ fixed. 

What do the errors on $\M$ and $\delta$ imply for the measurability of individual component masses $m_1$, $m_2$?
For $m_1 \leq m_2$ one has
\be
m_{1,2}(\M,\delta) = 2^{1/5}(1 \mp \delta)(1-\delta^2)^{-3/5}\,\M.
\ee
The error propagation formula then yields
\begin{eqnarray}
\left(\frac{\Delta m_i}{m_i}\right)^2 &=& \left(\frac{\partial \ln m_i}{\partial \ln\M}\right)^2\left(\frac{\Delta \M}{\M}\right)^2 + \left(\frac{\partial \ln m_i}{\partial \delta}\right)^2 (\Delta \delta)^2 
+ 2\,\left(\frac{\partial \ln m_i}{\partial \ln\M}\right)\left(\frac{\partial \ln m_i}{\partial \delta}\right)\,\Sigma^{\ln(\M),\delta}\nonumber\\
&=& \left(\frac{\Delta \M}{\M}\right)^2 + \left(\frac{5-\delta}{5(1-\delta^2)}\right)^2 (\Delta \delta)^2 
+ 2\,\left(\frac{\mp 5 + \delta}{5(1-\delta^2)}\right)\,\sigma^{\ln(\M),\delta} \frac{\Delta\M}{\M} \Delta\delta
\label{masserrors}
\end{eqnarray}
for $i = 1,2$; in the last term of the RHS the upper sign refers to $i=1$.
With restricted waveforms these tend to be very large \cite{CF}. The situation is dramatically different if amplitude corrections are taken into account, as can be seen in Fig.~\ref{compmassesfixedq}. In Advanced LIGO, for $q_m = 0.1$ systems at 100 Mpc, with amplitude-corrected waveforms the relative error $\Delta m_1/m_1$ (resp.~$\Delta m_2/m_2$) remains smaller than 10\% (resp.~15\%) for $22\,M_\odot \lesssim M \lesssim 200\,M_\odot$, while with the restricted waveform these errors reach 100\% already at $M \simeq 50\,M_\odot$. In EGO, with amplitude corrections $\Delta m_1/m_1$ (resp.~$\Delta m_2/m_2$) is smaller than 2\% (resp.~4\%) for total mass up to $500\,M_\odot$!

For a fixed total mass of $M=100\,M_\odot$, the relative errors $\Delta m_i/m_i$, $i=1,2$, vary significantly with mass ratio $q_m$ (Fig.~\ref{compmassesfixedM}), getting smaller as the system becomes more symmetric. For the given total mass, increasing $q_m$ from 0.1 to 1 makes a difference in the errors of more than a factor of three in both detectors, bringing the errors down to just a few percent in Advanced LIGO and a few tenths of a percent in EGO. Note that the errors do not perfectly coincide at $q_m=1$. In the last term in the right hand side of Eq.~(\ref{masserrors}), the correlation $\sigma^{\ln(\mathcal{M}),\delta}$ does not become zero at $\delta = 0$, and $[\partial \ln m_1/\partial \delta]_{\delta=0} = -[\partial \ln m_2/\partial \delta]_{\delta=0}$ while $[\partial \ln m_1/\partial \ln \mathcal{M}]_{\delta=0} = [\partial \ln m_2/\partial \ln \mathcal{M}]_{\delta=0}$, which is how the difference in errors comes about. This is just an artefact of the quadratic approximation used here to compute errors; in the right hand side of (\ref{masserrors}) all contributions beyond second order are ignored.

We end with the spin-related parameters. Again we start with the case of fixed $q_m$ but varying $M$. As before the spins are zero so that $\beta=\sigma=0$. With Advanced LIGO the spin-orbit parameter $\beta$ (Fig.~\ref{spinsFigure}) is reasonably well determined with restricted waveforms for lighter systems, but it quickly becomes effectively unmeasurable as total mass is increased. The amplitude-corrected waveforms do quite well up to $M \sim 150\,M_\odot$. In EGO, the restricted waveforms once again lead to reasonably small errors at the low mass end but a quick deterioration is seen with increasing $M$. The amplitude-corrected waveforms lead to $\Delta\beta \lesssim 0.6$ throughout the given mass range. This is to be compared with the largest value $|\beta|$ can take ($\simeq$ 9.4) if the binary's components are sub-extremal spinning black holes. Finally, in Advanced LIGO, both with restricted and amplitude-corrected waveforms one has $\Delta\sigma \gg 2.5$, i.e.~much larger than the largest value $\sigma$ may be expected to take for sub-extremal black holes, which is why we did not include a figure. Also in EGO the spin-spin parameter is poorly determined even with amplitude-corrected waveforms. This may well be related to the observation by Arun et al.~\cite{Testing3} that if the phasing coefficients $\psi_i$ are formally treated as independent parameters, the one at 2PN turns out to be by far the most poorly determined; recall that this is the order at which $\sigma$ appears.

For fixed total mass, the dependence of $\Delta\beta$ and $\Delta\sigma$ on mass ratio is fairly weak, as illustrated by Fig.~\ref{spinsFigureq}.

\begin{table}[htp!] 
\caption{Change in signal-to-noise ratio and improvement of parameter estimation for Advanced LIGO and EGO with increasing $p$ in $(p,2.5)$PN waveforms.}
\begin{tabular}{lrcccccc} 
\hline 
\hline 
AdvLIGO, $(5,50)\,M_\odot$ &&&&&&&\\
\hline
$p$ &\vline& \,\,\,SNR\,\,\, & \,\,\,$\Delta\M/\M$ (\%)\,\,\, & \,\,\,$\Delta\delta$ (\%)\,\,\, & \,\,\,$\Delta t_c$ (ms)\,\,\, & \,\,\,$\Delta\beta$\,\,\, & \,\,\,$\Delta\sigma$\,\,\,
\\
\hline
0   &\vline& 76.5 & 13.75 & 514.4 & 234.3 & 51.93 & 790.1 \\
0.5 &\vline& 85.0 & 0.905 & 4.498 & 6.468 & 2.143 & 8.529 \\ 
1   &\vline& 74.1 & 0.674 & 3.662 & 5.446 & 1.691 & 6.908 \\
1.5 &\vline& 69.0 & 0.463 & 4.376 & 5.273 & 1.124 & 6.067 \\
2   &\vline& 65.5 & 0.458 & 3.444 & 4.205 & 1.144 & 5.703 \\
2.5 &\vline& 64.0 & 0.471 & 2.318 & 3.822 & 1.611 & 4.457 \\
\hline
\hline
EGO, $(10,100)\,M_\odot$ &&&&&&&\\ 
\hline
$p$ &\vline& \,\,\,SNR\,\,\, & \,\,\,$\Delta\M/\M$ (\%)\,\,\, & \,\,\,$\Delta\delta$ (\%)\,\,\, & \,\,\,$\Delta t_c$ (ms)\,\,\, & \,\,\,$\Delta\beta$\,\,\, & \,\,\,$\Delta\sigma$\,\,\,
\\
\hline
0   &\vline& 461.9 & 2.140 & 81.75 & 74.91 & 8.160 & 125.4 \\
0.5 &\vline& 513.0 & 0.145 & 0.758 & 2.085 & 0.347 & 1.398 \\
1   &\vline& 446.4 & 0.107 & 0.625 & 1.769 & 0.272 & 1.145 \\
1.5 &\vline& 417.7 & 0.075 & 0.748 & 1.753 & 0.184 & 1.029 \\
2   &\vline& 396.8 & 0.076 & 0.589 & 1.405 & 0.194 & 0.970 \\
2.5 &\vline& 387.7 & 0.076 & 0.401 & 1.262 & 0.192 & 0.755 \\
\hline
\end{tabular} 
\label{t:SNR}
\end{table}

\subsection{Dependence on PN order of the amplitudes}

Table \ref{t:SNR} shows how errors on $(\ln(\M), \delta, t_c, \beta, \sigma)$ vary with $p$ for systems with $q_m=0.1$ and masses for which the SNRs of both the restricted and amplitude-corrected waveforms are close to maximum (specifically, $(5,50)\,M_\odot$ in Advanced LIGO and $(10,100)\,M_\odot$ in EGO). The restricted, $p=0$, waveform contains only the Newtonian harmonic at twice the orbital phase, and the independent parameters characterizing it are $(\ln(\mathcal{A}),\ln(\M),\delta,t_c,\psi_c,\beta,\sigma)$.
As explained near the end of subsection \ref{ss:waveforms}, for $p=0.5$ one has two extra harmonics and two more parameters become measurable in principle: $\cos(\iota)$ and one of the variables $(\cos(\theta),\phi,\psi)$, although in practice only $\cos(\iota)$ is measurable with any degree of accuracy. Each subsequent half-integer increase in $p$ leads to an additional harmonic. However, even for $p=2.5$ the parameters $(\cos(\theta),\phi,\psi)$ are poorly determined and extremely strongly correlated, as seen in Eqs.~(\ref{correlations}), and this is true \emph{a fortiori} for $p < 2.5$.

For those parameters that are measurable, the biggest improvement occurs when going from $p=0$ to $p=0.5$. At $p=2.5$ the errors tend to be about a factor of two smaller than at $p=0.5$. We note that from $p=0.5$ onward, all errors generally tend to decrease (with minor fluctuations), despite the fact that SNRs decrease monotonically. If we had presented errors for fixed SNR, one would have seen an even more pronounced improvement in parameter estimation. The great improvement seen in going from $p=0$ to $p=0.5$ can not be accounted for by this being the only case where SNR increases; if we had compared these cases for equal SNR then the improvement would still have been very large. Thus, it is the extra structure in the additional harmonics that is responsible for the improvement, as anticipated in subsection \ref{ss:structure}.

For the given system in Advanced LIGO, with the restricted waveform all of the intrinsic parameters are unmeasurable (except for $\M$, but still with a large error), while with the exception of $\sigma$ they are reasonably well determined using the $(2.5,2.5)$PN waveform. In the case of EGO, note how $\beta$ appears to be unmeasurable with the restricted waveform while the one at $(2.5,2.5)$PN does very well; to a certain extent the same is true of $\sigma$.

\begin{table}[htp!] 
\caption{Errors for different magnitudes and orientations of the spins with respect to $\hat{L} = (0,0,1)$.}
\begin{tabular}{ccrccrccccc} 
\hline 
\hline 
AdvLIGO,    & $(5,50)\,M_\odot$ &&&&&&& \\ 
\hline
$\quad \bar{\chi}_1 \quad$ & $\quad \bar{\chi_2} \quad$ & \vline & $\quad \beta \quad$ & $\quad \sigma \quad$ &\vline&  \,\,\,$\Delta\M/\M$ (\%)\,\,\, & \,\,\,$\Delta\delta$ (\%)\,\,\, & \,\,\,$\Delta t_c$ (ms)\,\,\, & $\quad \Delta \beta \quad$  & $\quad \Delta \sigma \quad$\\
\hline
$(0,0,0.9)$     & $(0,0,0.9)$     &\vline& 8.004  & 0.661  &\vline& 0.449 & 2.262 & 3.489 & 1.314 & 4.296 \\
$(0,0,0.9)$     & $(0,0,-0.9)$    &\vline& -6.934 & -0.661 &\vline& 0.494 & 2.314 & 4.031 & 1.214 & 4.677 \\
$(0.67,0,0.67)$ & $(0,0.67,0.67)$ &\vline& 5.958  & 0.366  &\vline& 0.456 & 2.317 & 3.581 & 1.269 & 4.383 \\
\hline
$(0,0,1.5)$     & $(0,0,1.5)$     &\vline& 13.340  & 1.836  &\vline& 0.435 & 2.201 & 3.350 & 1.477 & 4.190 \\
$(0,0,1.5)$     & $(0,0,-1.5)$    &\vline& -11.557 & -1.836 &\vline& 0.529 & 2.309 & 4.253 & 1.338 & 4.853 \\
$(0.67,0,0.67)$ & $(0,1.4,1.4)$   &\vline& 12.017  & 0.766  &\vline& 0.438 & 2.192 & 3.366 & 1.428 & 4.139 \\
\hline 
\hline
EGO,        & $(10,100)\,M_\odot$ &&&&&&& \\
\hline
$\quad \bar{\chi}_1 \quad$ & $\quad \bar{\chi_2} \quad$ & \vline & $\quad \beta \quad$ & $\quad \sigma \quad$ &\vline&  \,\,\,$\Delta\M/\M$ (\%)\,\,\, & \,\,\,$\Delta\delta$ (\%)\,\,\, & \,\,\,$\Delta t_c$ (ms)\,\,\, & $\quad \Delta \beta \quad$  & $\quad \Delta \sigma \quad$ \\
\hline
$(0,0,0.9)$     & $(0,0,0.9)$     &\vline& 8.004  & 0.661  &\vline& 0.073 & 0.387 & 1.170 & 0.220 & 0.732 \\
$(0,0,0.9)$     & $(0,0,-0.9)$    &\vline& -6.934 & -0.661 &\vline& 0.080 & 0.391 & 1.356 & 0.198 & 0.786 \\
$(0.67,0,0.67)$ & $(0,0.67,0.67)$ &\vline& 5.958  & 0.366  &\vline& 0.074 & 0.396 & 1.199 & 0.211 & 0.746 \\
\hline
$(0,0,1.1)$     & $(0,0,1.1)$    &\vline& 9.783  & 0.987   &\vline& 0.072 & 0.385 & 1.159 & 0.227 & 0.718 \\
$(0,0,1.1)$     & $(0,0,-1.1)$   &\vline& -8.475 & -0.987  &\vline& 0.080 & 0.419 & 1.355 & 0.214 & 0.753 \\
$(0.67,0,0.67)$ & $(0,1.1,1.1)$  &\vline& 9.527  & 0.601   &\vline& 0.072 & 0.385 & 1.158 & 0.225 & 0.714 \\
\hline
\end{tabular} 
\label{t:Spin}
\end{table}

\subsection{Dependence on the magnitudes and orientations of the spins}
\label{ss:spindependence}

So far we have restricted ourselves to systems with zero spins; although $\beta$ and $\sigma$ were included as variables in the error analysis and  the errors $\Delta\beta$ and $\Delta\sigma$ were investigated, the \emph{values} of $\beta$ and $\sigma$ were set to zero (i.e., the Fisher matrix was evaluated on the hypersurface $\beta=\sigma=0$) until now. In Table \ref{t:Spin} we look at the errors for different values of the spin-related quantities. The focus is on binary systems where the spins of the component objects are close to the Kerr bound, both with $||\bar{\chi}_i|| < 1$ and $||\bar{\chi}_i|| > 1$, $i=1,2$. Objects in the latter category could be Kerr black holes that exceed extremality, boson stars \cite{bosonstars}, or other exotica. The spins are chosen at various angles with respect to the orbital angular momentum vector. By looking at the results of Table \ref{t:Spin}, and also comparing with those of Table \ref{t:SNR} for $p=2.5$, one sees that the errors have but a weak dependence on the values of $\beta$ and $\sigma$. Note that $\Delta\mathcal{M}/\mathcal{M}$ and $\Delta t_c$ tend to be somewhat smaller if spins are aligned and larger when they are anti-aligned.

The parameters $\beta$ and $\sigma$ are essentially unmeasurable with restricted waveforms. By contrast, with $(2.5,2.5)$PN waveforms, in Advanced LIGO and EGO one can measure $\beta$; the parameter $\sigma$ remains poorly determined even in EGO. The accuracy with which $\beta$ can be measured would already allow for the detection of even relatively small violations of the Kerr bound. Apart from spins, the expression (\ref{beta}) for $\beta$ depends only on the relative mass difference $\delta$, which is rather well determined. Now, if $|\beta|$ is larger than the largest value allowed by $\delta$ assuming sub-extremal black holes, then at least one component of the binary must have a spin exceeding the Kerr bound. For any given value of $\delta$, this ``maximum value'' $\beta_{max}(\delta)$ corresponds to the case where $\bar{\chi}_i\cdot\hat{L}=1$, $i=1,2$, and one has 
\be
\beta_{max}(\delta) = \frac{1}{12}\left[113-\frac{76}{4}(1-\delta^2)\right].
\label{betamax}
\ee
We would like to know how well one can measure the difference $\mathcal{B}(\beta,\delta) \equiv |\beta| - \beta_{max}(\delta)$. The error propagation formula yields
\begin{eqnarray}
(\Delta\mathcal{B})^2 &=& (\Delta\beta)^2 + \left(\frac{\partial \beta_{max}}{\partial \delta}\right)^2(\Delta\delta)^2
\mp 2 \left(\frac{\partial \beta_{max}}{\partial \delta}\right)\,\Sigma^{\beta,\delta} \nonumber\\
&=&  (\Delta\beta)^2 + \frac{361}{36}\delta^2 (\Delta \delta)^2 \mp \frac{19}{3}\delta\,\sigma^{\beta,\delta}\Delta\beta\Delta\delta,
\label{DeltaB}
\end{eqnarray}
where in the last term, the upper sign is for the case $\beta > 0$ and the lower one for $\beta < 0$. At $\beta=0$ the two expressions for $\Delta\mathcal{B}$ disagree. This discrepancy is analogous to the one encountered in subsection \ref{ss:intrinsic} for $\Delta m_i/m_i$, $i=1,2$, at $\delta=0$; again the error is only given in the leading order, quadratic approximation. In any case, the last term in (\ref{DeltaB}) tends to be small compared to the other terms, including at $\beta = 0$.

Let us discuss a few examples to probe the extent to which violations of the Kerr bound can be measured in Advanced LIGO and EGO. In the above expression for $\Delta\mathcal{B}$, $\delta$ refers to the measured value of that parameter; for convenience we set it equal to its physical value. Note that in order to claim detection of a Kerr bound violation, it is not necessary for $\mathcal{B}$ itself to be measurable with great accuracy; all that is needed is the ability to claim $\mathcal{B} > 0$ with a reasonable degree of confidence. We take as a benchmark that $2\Delta\mathcal{B}$ be no larger than $\mathcal{B}$.

First let the detector be Advanced LIGO and consider a $(5,50)\,M_\odot$ system (so that $\delta = 9/11$) with $\bar{\chi}_1=\bar{\chi}_2= 1.5\,\hat{L}$. Then one has $2\Delta \mathcal{B} = 3.032$. This is to be compared with $\mathcal{B} = 4.447$ computed from the actual values of $\beta$ and $\delta$. By the above criterion, a violation of the Kerr bound of this size would be detectable.

Still considering Advanced LIGO and the same component masses, now let $\bar{\chi}_1=(0.67,0,0.67)$ and $\bar{\chi}_2 = (0,1.4,1.4)$, setting $\hat{L} = (0,0,1)$. In this case the first object is sub-extremal ($||\bar{\chi}_1|| \simeq 0.95$) while the other exceeds the Kerr bound with $||\bar{\chi}_2|| \simeq 1.98$. One has $2\Delta \mathcal{B} = 2.930$ while $\mathcal{B} = 3.123$. Hence also in this case the violation of the Kerr bound can be detected. As this example illustrates, if the spins are not aligned with each other and $\hat{L}$ (as will generally be the case), then the Kerr bound will need to be exceeded more strongly for the effect to be observable. 

In EGO the parameter $\beta$ is much better determined. Consider a $(10,100)\,M_\odot$ system, this time with $\bar{\chi}_1 = \bar{\chi}_2 = 1.1\,\hat{L}$. Here one has $2\Delta \mathcal{B} = 0.466$ while $\mathcal{B} = 0.889$; hence a violation of the Kerr bound at this level can be detected with EGO.

Finally, take a $(10,100)\,M_\odot$ system in EGO with $\bar{\chi}_1 = (0.67,0,0.67)$ and $\bar{\chi}_2=(0,1.1,1.1)$ for $\hat{L} = (0,0,1)$, so that one object is sub-extremal with $||\bar{\chi}_1|| \simeq 0.95$ while the other exceeds the Kerr bound with $||\bar{\chi}_2|| \simeq 1.56$. Then $2\Delta\mathcal{B} = 0.462$ while $\mathcal{B} = 0.634$, allowing for detection of the Kerr bound violation. Once again, with non-aligned spins a stronger excess is required for it to be observable.

\subsection{A note on initial detectors}

Given the above results, it is of interest to similarly compare parameter estimation with restricted and amplitude-corrected waveforms in initial detectors, such as Initial LIGO. An important caveat is that here the SNRs are expected to be low, in which case the covariance matrix formalism may significantly underestimate the errors \cite{MonteCarlo}. We give results for two asymmetric systems\footnote{As discussed in subsection \ref{ss:parameterestimation}, the Fisher matrix for the restricted waveform becomes ill-conditioned when $m_1/m_2 \simeq 1$ because of the use of $\delta$ as a coordinate, while the Fisher matrix for the full waveform becomes singular at $m_1=m_2$ if $\ln(\eta)$ is used in place of $\delta$. For this reason we do not consider the equal mass case.} with masses $(1.4,10)\,M_\odot$ and $(5,15)\,M_\odot$, at a distance of 20 Mpc; spins are set to zero. Table \ref{t:initial} shows errors obtained with the restricted and full waveforms for these systems.
Again the amplitude-corrected waveform has significant advantages over the restricted one. Note that in the case of the lighter system, the full waveform allows for determination of $\beta$, at least at the given distance; $\sigma$ is unmeasurable.

\begin{table}[htp!] 
\caption{Parameter estimation with Initial LIGO. Distance is 20 Mpc; spins are zero.}
\begin{tabular}{lrcccccc} 
\hline 
\hline
$(1.4,10)\,M_\odot$ &&&&&&& \\
\hline
Waveform &\vline& \,\,\,SNR\,\,\, & \,\,\,$\Delta\M/\M$ (\%)\,\,\, & \,\,\,$\Delta\delta$ (\%)\,\,\, & \,\,\,$\Delta t_c$ (ms)\,\,\, & \,\,\,$\Delta\beta$\,\,\, & \,\,\,$\Delta\sigma$\,\,\,
\\
\hline
(0,2.5)PN    &\vline& 13.49 & 1.432 & 630.0 & 47.47 & 12.45 & 609.0 \\
(2.5,2.5)PN  &\vline& 11.99 & 0.262 & 29.54 & 2.542 & 1.223 & 29.17 \\
\hline
\hline
$(5,15)\,M_\odot$ &&&&&&& \\
\hline
Waveform &\vline& \,\,\,SNR\,\,\, & \,\,\,$\Delta\M/\M$ (\%)\,\,\, & \,\,\,$\Delta\delta$ (\%)\,\,\, & \,\,\,$\Delta t_c$ (ms)\,\,\, & \,\,\,$\Delta\beta$\,\,\, & \,\,\,$\Delta\sigma$\,\,\,
\\
\hline
(0,2.5)PN    &\vline& 26.78 & 24.72 & 4309.1 & 101.3 & 44.13 & 1671.3 \\
(2.5,2.5)PN  &\vline& 22.63 & 2.598 & 19.34  & 4.843 & 7.590 & 20.97 \\
\hline
\end{tabular} 
\label{t:initial}
\end{table}

\section{Summary and conclusions}
\label{s:Conclusions}

We have compared parameter estimation with restricted versus amplitude-corrected post-Newtonian gravitational waveforms, in the stationary phase approximation, for compact binary inspiral in Advanced LIGO and a possible third-generation ground-based detector called EGO. For the amplitude-corrected waveforms both phasing and amplitudes were considered up to 2.5PN order, with inclusion in the phasing of spin-orbit effects at 1.5PN and 2PN, and spin-spin effects at 2PN. In the case of restricted waveforms, the parameters of interest can be extracted only from the phase. With amplitude-corrected waveforms a wealth of information is carried by the spectrum as well, which leads to very significant improvements in parameter estimation. Because we mostly dealt with interferometers that will detect sources with high signal-to-noise ratios, it was justified to use the covariance method for estimating errors and this is what was done throughout. The amplitude-corrected waveforms considered here depend on 11 parameters, which makes it difficult to numerically invert the Fisher matrix. Fortunately the angles $(\theta,\phi,\psi)$ tend to have only small correlations with the other parameters, so that it is justified to work with a much better conditioned $8\,\times\,8$ Fisher matrix in which these angles are disregarded.

Our main results may be summarized as follows.

\begin{itemize}
\item The parameters that can be measured with reasonable precision using \emph{restricted} waveforms are 
\be
(\ln(\mathcal{A}),\ln(\M),\delta,t_c,\beta,\sigma), \label{measurable0}
\ee
although the logarithm of the amplitude, $\ln(\mathcal{A})$, does not carry much useful information. With restricted waveforms the errors on (\ref{measurable0}) -- not counting $\ln(\mathcal{A})$ -- are steep functions of total mass, and parameters quickly become unmeasurable as one moves towards the higher-mass end of stellar mass binaries.

Inclusion of amplitude corrections alters the situation dramatically. In that case the parameters that can be measured with good accuracy are
\be
(\ln(\M),\delta,t_c,\cos(\iota),\beta,\sigma). \label{measurable}
\ee
Here the errors increase much more slowly as total mass goes up. For sources at a distance of 100 Mpc, with amplitude-corrected waveforms parameter estimation remains possible for intermediate-mass binaries with total mass up to at least $250\,M_\odot$ in Advanced LIGO and $500\,M_\odot$ in EGO. Up to these masses, the relative error on chirp mass varies the most (by more than an order of magnitude in Advanced LIGO and two orders of magnitude in EGO); the errors on the other parameters increase by a factor of a few. We also investigated how errors improve with the PN order in amplitude; the largest improvement occurs in going from 0PN to 0.5PN, although apart from fluctuations, errors keep decreasing significantly as the order is increased further. At 2.5PN in amplitude, the errors have a relatively small dependence on the values of $\beta$ and $\sigma$; $\Delta\mathcal{M}/\mathcal{M}$ and $\Delta t_c$ tend to be a little smaller if spins are aligned and a little larger when they are anti-aligned.

\item Because our analysis was confined to single detectors, among the intrinsic parameters only the time of coalescence can be measured with good accuracy. (The inclination angle is also reasonably well determined, but it is not very interesting by itself.) At the low-mass end, the full waveform reduces the error on $t_c$  by an order of magnitude in Advanced LIGO and about a factor of 5 in EGO compared to the restricted waveform, but the differences become much larger as one moves to higher masses.

\item Using the restricted waveforms, component masses tend to be extremely poorly determined even for low-mass systems. This is no longer the case with amplitude-corrected waveforms: In Advanced LIGO the relative errors on $m_1$, $m_2$ already stay below 15\% in a significant part of the mass range, while in EGO they are below 4\% all the way up to a total mass of $500\,M_\odot$! These numbers refer to asymmetric systems; for equal component masses the relative errors will be smaller still. The best values seen here are as low as a few percent in Advanced LIGO and a few tenths of a percent in EGO. All this will have important astrophysical consequences: Unlike what the restricted PN approximation suggests, it will be possible to use binary inspiral events to perform black hole population studies, not just for stellar mass black holes but also for intermediate-mass black holes that would be the results of successive mergers.

\item The spin-spin parameter $\sigma$ occuring at 2PN in the phasing remains poorly determined even when the full waveform is used. With restricted waveforms, the spin-orbit parameter $\beta$ at 1.5PN is undetermined in Advanced LIGO; inclusion of amplitude corrections makes it measurable. In EGO $\beta$ is measurable with restricted waveforms for systems with low masses, but once again the amplitude-corrected waveforms lead to a very significant improvement. On the basis of this we have suggested a way to find out if one or both of the component objects violate the Kerr bound. With the full waveforms, the measured value of $|\beta|$ can be compared with the largest possible value this parameter can take given the measured relative mass difference and assuming that the binary consisted of two sub-extremal black holes. If for a particular inspiral event, $|\beta|$ is found to be larger than this maximum value, then at least one component object must be more exotic. In this way one would be able to detect fairly small violations of the Kerr bound, both with Advanced LIGO and EGO.

\item We also briefly considered parameter estimation in Initial LIGO, showing that there too the amplitude-corrected waveform can have a very significant advantage over the restricted one. 

\end{itemize}

Of necessity our analysis was somewhat limited, as the way in which spins enter the amplitudes has not been the subject of much attention (see, however, \cite{OTO}). The drastic improvements in the estimation of the other parameters due to sub-dominant harmonics suggest that knowledge of spin-related contributions to the amplitudes would hold significant advantages. Moreover, in the present work $\beta$ and $\sigma$ were considered constant. As shown by Vecchio \cite{Vecchio} and Lang and Hughes \cite{LangHughes} in the context of \emph{restricted} waveforms for LISA, taking into account the precession-induced modulation due to time-dependent spins also leads to qualitative improvements in parameter estimation.

In both the present paper and its precursor \cite{SNRpaper} we chose to restrict ourselves to ground-based detectors, but it should not be difficult to extend our work to the case of LISA. Preliminary studies in this direction have already been performed by Sintes and Vecchio \cite{SV2} and Hellings and Moore \cite{Hellings1,Hellings2}; in view of the results presented here it would be of great interest to perform a systematic and comprehensive investigation.

We end with a comment regarding \emph{initial} detectors. Although in the immediate future it would not be computationally feasible to use the full waveforms for inspiral signal searches, there would be clear advantages in using them for follow-up studies once a detection has been made. As indicated by both our own results and those of R\"over et al.~\cite{Roveretal}, this issue deserves much more attention than it has received hitherto.

\section*{Acknowledgements}

It is a pleasure to thank K.G.~Arun, S.~Babak, and B.S.~Sathyaprakash for very useful discussions. We are indebted to M.~Punturo for sharing with us the power spectral density for EGO. In addition we would like to thank the two anonymous referees whose comments greatly improved both the contents and the presentation of this paper. This research was supported by PPARC grant PP/B500731/1.

\section*{Appendix A. Second and third generation interferometric observatories}

The noise power spectral density (PSD) for Advanced LIGO is approximately \cite{Arunetal}
\be
S_h(f) = S_0\,\left[x^{-4.14}-5\,x^{-2}+111\frac{1-x^2+x^4/2}{1+x^2/2}\right],
\label{AdvLIGO_PSD}
\ee
where $x=f/f_0$, $f_0=215$ Hz, $S_0=10^{-49}\,\mbox{Hz}^{-1}$, and $f_s=20$ Hz. The low-frequency cut-off is taken to be $f_s = 20$ Hz; below this frequency $S_h(f)$ can be considered infinite for all practical purposes. 

\begin{figure}[htbp!] 
\includegraphics[scale=0.5,angle=0]{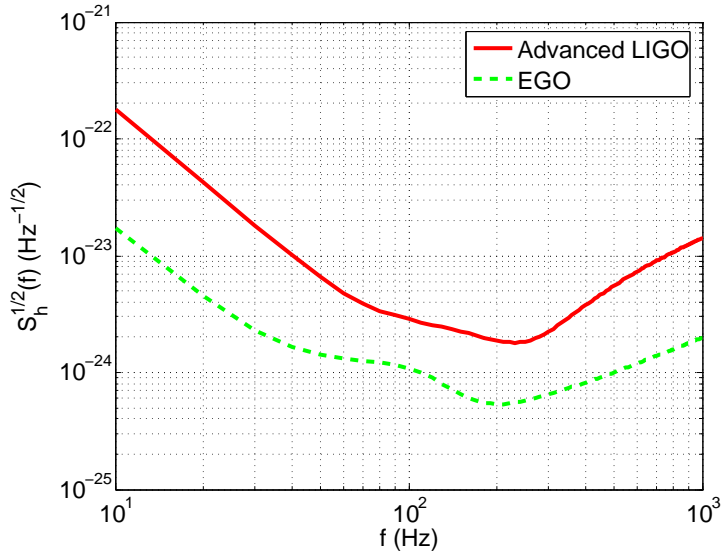}
\caption{The strain sensitivities for Advanced LIGO and EGO.}
\label{f:PSDs}
\end{figure}

The basic technical specifications of EGO were outlined in \cite{SNRpaper}, where an analytic fit of its PSD was also introduced. For completeness we repeat the expression for the PSD. Above frequencies of a few Hertz it is well-approximated by
\be
S_h(f) = S_0\,\left[x^{p_1} + a_1 x^{p_2} + a_2 \frac{1+b_1 x+b_2 x^2 + b_3 x^3 + b_4 x^4 + b_5 x^5 + b_6 x^6}{1 + c_1 x + c_2 x^2 + c_3 x^3 + c_4 x^4}\right]
\ee
where again $x = f/f_0$, this time with $f_0=200$ Hz, and $S_0 = 1.61 \times 10^{-51}\,\mbox{Hz}^{-1}$. The low-frequency cut-off is $f_s = 10$ Hz. One has
\ba
p_1 = -4.05, &\quad& p_2 = -0.69,  \nn\\
a_1 = 185.62, &\quad& a_2 = 232.56, \nn\\
b_1 = 31.18, \quad b_2 = -64.72, \quad b_3 = 52.24, &\quad& b_4 = -42.16, \quad b_5 = 10.17, \quad b_6 = 11.53, \nn\\
c_1 = 13.58, \quad c_2 = -36.46,  &\quad& c_3 = 18.56, \quad c_4 = 27.43. \nn\\
\ea
The strain sensitivities for Advanced LIGO and EGO are plotted in Fig.~\ref{f:PSDs}.


\begin{thebibliography}{99}

\bibitem{Grishchuketal} L.P.~Grishchuk, V.M.~Lipunov, K.A.~Postnov, M.E.~Prokhorov, and B.S.~Sathyaprakash,
Phys.~Usp.~{\bf 44}, 1-51 (2001);\\
L.P.~Grishchuk, V.M.~Lipunov, K.A.~Postnov, M.E.~Prokhorov, and B.S.~Sathyaprakash,
Usp.~Fiz.~Nauk.~{\bf 171}, 3-59 (2001);\\
L.P.~Grishchuk,
in \emph{Astrophysics Update}, ed.~J.W.~Mason (Springer-Praxis, Berlin, 2004)

\bibitem{Observatories} http://www.ligo.caltech.edu/; http://www.virgo.infn.it/;
http://www.geo600.uni-hannover.de/; http://tamago.mtk.nao.ac.jp/
\bibitem{Peters} P.C.~Peters, 
Gravitational Radiation and the Motion of Two Point Masses,
Phys.~Rev.~{\bf 136}, B1224-1232 (1964)

\bibitem{Blanchet} L.~Blanchet, 
Liv.~Rev.~Rel.~{\bf 5}, 3 (2002)

\bibitem{Testing1} E.~Berti, A.~Buonanno, and C.M.~Will,
Phys.~Rev.~D {\bf 71}, 084025 (2005)
\bibitem{Testing2} E.~Berti, A.~Buonanno, and C.M.~Will,
Class.~Quantum Grav.~{\bf 22}, S943-S954 (2005)

\bibitem{Testing3} K.G.~Arun, B.R.~Iyer, M.S.S.~Qusailah, and B.S.~Sathyaprakash,
Class.~Quantum Grav.~{\bf 23}, L37-L43 (2006);\\
K.G.~Arun, B.R.~Iyer, M.S.S.~Qusailah, and B.S.~Sathyaprakash, 
gr-qc/0604067

\bibitem{2.5PN} 
L.~Blanchet, B.R.~Iyer, C.M.~Will, and A.G.~Wiseman,
Class.~Quantum Grav.~{\bf 13}, 575-584 (1996);\\
K.G.~Arun, L.~Blanchet, B.R.~Iyer, and M.S.S.~Qusailah,
Class.~Quantum Grav.~{\bf 21}, 3771-3802 (2004);
Erratum ibid.~{\bf 22}, 3115 (2005) 

\bibitem{3.5PN}
T.~Damour, P.~Jaranowski, and G.~Sch\"afer,
Phys.~Lett. B {\bf 513}, 147 (2001);\\
Y.~Itoh, T.~Futamase, and H.~Asada,
Phys.~Rev.~D {\bf 63}, 064038 (2001);\\
L.~Blanchet, G.~Faye, B.R.~Iyer, and B.~Joguet,
Phys.~Rev.~D {\bf 65}, 061501(R) (2002);
Erratum ibid.~D {\bf 71}, 129902 (2005);\\
Y.~Itoh and T.~Futamase, 
Phys.~Rev.~D {\bf 68}, 121501(R) (2003);\\
Y.~Itoh, 
Phys.~Rev.~D {\bf 69}, 064018 (2004);\\
L.~Blanchet, T.~Damour, and G.~Esposito-Far\`ese,
Phys.~Rev.~D {\bf 69}, 124007 (2004);\\
L.~Blanchet, T.~Damour, G.~Esposito-Far\`ese, and B.R.~Iyer,
Phys.~Rev.~Lett.~{\bf 93}, 091101 (2004);\\
Y.~Itoh, 
Class.~Quantum Grav.~{\bf 21}, S529-S534 (2004);\\
L.~Blanchet and B.R.~Iyer,
Phys.~Rev.~D {\bf 71}, 024004 (2005)

\bibitem{Helstrom} C.W.~Helstrom, \emph{Statistical Theory of Signal Detection} (Pergamon Press,
Cambridge, England, 1968);\\
S.~Babak, R.~Balasubramanian, D.~Churches, T.~Cokelaer, and B.S.~Sathyaprakash, gr-qc/0604037;\\
S.~Babak, T.~Cokelaer, B.S.~Sathyaprakash, and A.S.~Sengupta, \emph{in preparation}

\bibitem{last3minutes} C.~Cutler et al.,
Phys.~Rev.~Lett.~{\bf 70}, 2984-2987 (1993)

\bibitem{Letter} C.~Van Den Broeck,
Class.~Quantum Grav.~{\bf 23}, L51-L58 (2006)

\bibitem{SNRpaper} C.~Van Den Broeck and A.S.~Sengupta,
Class.~Quantum Grav.~{\bf 24}, 155-176 (2007)

\bibitem{ringdown} O.~Dreyer, B.~Kelly, B.~Krishnan, L.S.~Finn, D.~Garrison, and R.~Lopez-Aleman,
Class.~Quantum Grav.~{\bf 21}, 787-804 (2004);\\
E.~Berti, V.~Cardoso, and C.M.~Will,
Phys.~Rev.~D {\bf 73}, 064030 (2006)

\bibitem{SV1} A.M.~Sintes and A.~Vecchio,
in \emph{Proceedings of the Rencontres de Moriond: Gravitational waves and experimental gravity}, 
ed.~J.~Dumarchez, Editions Fronti\`eres, 2000

\bibitem{Roveretal} C.~R\"over, R.~Meyer, and N.~Christensen, gr-qc/0609131

\bibitem{SV2} A.M.~Sintes and A.~Vecchio,
in \emph{Proceedings of the Third Amaldi Conference on Gravitational Waves}, ed.~S.~Meshkov, American Institute of Physics, 2000

\bibitem{Hellings1} T.A.~Moore and R.W.~Hellings,
Phys.~Rev.~D {\bf 65}, 062001 (2002)

\bibitem{Hellings2} R.W.~Hellings and T.A.~Moore,
Class.~Quantum Grav.~{\bf 20}, S181-S192 (2003)

\bibitem{Punturo} M.~Punturo, private communication (2006)

\bibitem{spin} L.~Blanchet, T.~Damour, B.~Iyer, C.M.~Will, A.G.~Wiseman,
Phys.~Rev.~Lett.~{\bf 74}, 3515-3518 (1995)

\bibitem{spin2.5PN} G.~Faye, L.~Blanchet, and A.~Buonanno,
gr-qc/0605139;\\
L.~Blanchet, A.~Buonanno, and G.~Faye,
gr-qc/0605140

\bibitem{SPA} K.S.~Thorne,
in \emph{300 Years of Gravitation}, eds.~S.W.~Hawking and W.~Israel (Cambridge University Press, Cambridge, 
England, 1987), p.~330;\\
B.S.~Sathyaprakash and S.V.~Dhurandhar,
Phys.~Rev.~D {\bf 44}, 3819-3934 (1991)

\bibitem{Finn} L.S.~Finn, Phys.~Rev.~D {\bf 46}, 5236 (1992)

\bibitem{MonteCarlo} K.~D.~Kokkotas, A.~Kr\'olak, and G.~Tsegas,
Class.~Quantum Grav.~{\bf 11}, 1901 (1994);\\
R.~Balasubramanian, B.S.~Sathyaprakash, and S.V.~Dhurandhar,
Phys.~Rev.~D {\bf 53}, 3033 (1996);
Erratum ibid.~{\bf 54}, 1860 (1996);\\
R.~Balasubramanian and S.V.~Dhurandhar,
Phys.~Rev.~D {\bf 57}, 3408 (1998)

\bibitem{Arunetal} K.G.~Arun, B.R.~Iyer, B.S.~Sathyaprakash and P.A.~Sundararajan,
Phys.~Rev. D {\bf 71}, 084008 (2005); Erratum ibid.~{\bf 72}, 069903 (2005)

\bibitem{QuinlanShapiro} G.D.~Quinlan and S.L.~Shapiro, 
Astrophys.~J.~{\bf 343}, 725-749 (1989);\\
G.D.~Quinlan and S.L.~Shapiro,
Astrophys.~J.~{\bf 356}, 483-500 (1990);\\
K.~Gultekin, M.~Coleman Miller and D.P~Hamilton,
Astrophys.~J,~{\bf 616}, 221-230 (2004)

\bibitem{FFT1} S.~Droz, D.J.~Knapp, E.~Poisson, and B.J.~Owen,
Phys.~Rev.~D {\bf 59} 124016 (1999)

\bibitem{FFT2} T.~Damour, B.R.~Iyer, and B.S.~Sathyaprakash, 
Phys.~Rev.~D {\bf 62}, 084036 (2000)

\bibitem{KWW} L.E.~Kidder, C.M.~Will, A.G.~Wiseman,
Phys.~Rev.~D {\bf 47}, 4183-4187 (1993)

\bibitem{PoissonWill} E.~Poisson and C.M.~Will,
Phys.~Rev.~D {\bf 52}, 848-855 (1995)

\bibitem{Tinto} S.V.~Dhurandhar and M.~Tinto, Mon.~Not.~R.~Astron.~Soc.~{\bf 234}, 663-676 (1988);\\
S.V.~Dhurandhar and M.~Tinto, Mon.~Not.~R.~Astron.~Soc.~{\bf 236}, 621-627 (1989);\\
Y.~G\"ursel and M.~Tinto, in \emph{Proceedings of the 12th International Conference on General Relativity and Gravitation}, ed.~N.~Ashby (Cambridge University Press, Cambridge, 1989), p.~573

\bibitem{CF} C.~Cutler and \'E.\'E.~Flanagan,
Phys.~Rev.~D {\bf 49}, 2658-2697 (1994)

\bibitem{bosonstars} E.W.~Mielke and F.E.~Schunck,
in \emph{Proceedings of the 8th Marcel Grossmann Meeting}, ed.~T.~Piran (World Scientific, Singapore, 1998)

\bibitem{OTO} B.J.~Owen, H.~Tagoshi, and A.~Ohashi,
Phys.~Rev.~D {\bf 57}, 6168-6175 (1998)

\bibitem{Vecchio} A.~Vecchio,
Phys.~Rev.~D {\bf 70}, 042001 (1994)

\bibitem{LangHughes} R.N.~Lang and S.A.~Hughes,
gr-qc/0608062 


\end{thebibliography}
\end{document}